\documentclass[floatfix,preprint,amsmath,amssymb,onecolumn,superscriptaddress,showpacs,longbibliography,11pt]{revtex4-2}
\usepackage{amsmath,amssymb}
\usepackage{graphicx}
\usepackage{appendix}

\usepackage[usenames,dvipsnames]{color}
\usepackage[colorlinks=true, citecolor=blue, linkcolor=blue, urlcolor=blue]{hyperref}
\usepackage{bm}
\usepackage{cleveref} 
\usepackage{braket}
\usepackage[nottoc,numbib]{tocbibind}
\usepackage{mathtools}
\usepackage{layouts}
\usepackage{lineno}

\setcitestyle{super}

\newcommand{\ada}{a^{\dagger}a}
\newcommand{\cdc}{c^{\dagger}_k c_k}
\newcommand{\aplusa}{a^{\dagger} {+} a}

\AtBeginDocument{}

\begin{document}

\title{Quantum Floquet engineering with an exactly solvable tight-binding chain in a cavity}

\author{Christian J.~Eckhardt}
\thanks{These authors contributed equally}
\affiliation 
{Institut f\"ur Theorie der Statistischen Physik, RWTH Aachen University and JARA-Fundamentals of Future Information Technology, 52056 Aachen, Germany}
\affiliation 
{Max Planck Institute for the Structure and Dynamics of Matter, Center for Free-Electron Laser Science, Luruper Chaussee 149, 22761 Hamburg, Germany}

\author{Giacomo Passetti}
\thanks{These authors contributed equally}
\affiliation 
{Institut f\"ur Theorie der Statistischen Physik, RWTH Aachen University and JARA-Fundamentals of Future Information Technology, 52056 Aachen, Germany}

\author{Moustafa Othman}
\affiliation 
{Technische Universität Braunschweig, Institut für Mathematische Physik, Mendelssohnstraße 3, 38106 Braunschweig, Germany}

\author{Christoph Karrasch}
\affiliation 
{Technische Universität Braunschweig, Institut für Mathematische Physik, Mendelssohnstraße 3, 38106 Braunschweig, Germany}

\author{Fabio Cavaliere}
\affiliation
{Dipartimento di Fisica, Università di Genova, 16146, Genova, Italy}
\affiliation
{SPIN-CNR, 16146, Genova, Italy}

\author{Michael A.~Sentef}
\affiliation 
{Max Planck Institute for the Structure and Dynamics of Matter, Center for Free-Electron Laser Science, Luruper Chaussee 149, 22761 Hamburg, Germany}

\author{Dante M.~Kennes}
\email{dante.kennes@rwth-aachen.de}
\affiliation 
{Institut f\"ur Theorie der Statistischen Physik, RWTH Aachen University and JARA-Fundamentals of Future Information Technology, 52056 Aachen, Germany}
\affiliation 
{Max Planck Institute for the Structure and Dynamics of Matter, Center for Free-Electron Laser Science, Luruper Chaussee 149, 22761 Hamburg, Germany}

\date{\today}
\begin{abstract}

Recent experimental advances enable the manipulation of quantum matter by exploiting the quantum nature of light. However,
paradigmatic exactly solvable models, such as the Dicke, Rabi or Jaynes-Cummings models for quantum-optical systems, are scarce in the corresponding solid-state, quantum materials context.
Focusing on the long-wavelength limit for the light, here, we provide such an exactly solvable model given by a tight-binding chain coupled to a single cavity mode via a quantized version of the Peierls substitution.
We show that perturbative expansions in the light-matter coupling have to be taken with care and can easily lead to a false superradiant phase.
Furthermore, we provide an analytical expression for the groundstate in the thermodynamic limit, in which the cavity photons are squeezed by the light-matter coupling.
In addition, we derive analytical expressions for the electronic single-particle spectral function and optical conductivity. We unveil quantum Floquet engineering signatures in these dynamical response functions, such as analogs to dynamical localization and replica side bands,  complementing paradigmatic classical Floquet engineering results.
Strikingly, the Drude weight in the optical conductivity of the electrons is partially suppressed by the presence of a single cavity mode through an induced electron-electron interaction.

\end{abstract}
\maketitle

\section{Introduction}\label{INTRO}
The control of matter through light, or more generally electromagnetic (EM) radiation, is a research direction that has gained tremendous attention recently.\cite{colloquium}
It connects to many topical fields including information processing and steering chemical reactions.\cite{Ac_n_2018,Moody2021, Ebbesen2016, feist_polaritonic_2018, Ruggenthaler2018, ribeiro_polariton_2018, flick_strong_2018, FriskKockum2019}  
In recent years, some exciting progress has been made towards this goal by periodically driving materials with light in a regime where the quantum nature of the light field can be disregarded. \cite{wang_observation_2013, mciver_light-induced_2020}
 In this classical-light regime the physics of materials under continuous-wave irradiation is efficiently described by Floquet theory. \cite{bukov_universal_2015, Eckardt2017, oka_floquet_2019} Within Floquet theory, a time-periodic Hamiltonian is replaced by a quasi-static, effective so-called Floquet Hamiltonian, which can include renormalized effective model parameters, new synthetically generated terms, as well as Floquet sidebands, i.e., shakeoff features separated by the driving frequency from the main resonances, in frequency-dependent spectra.
The search for driving protocols that realize certain effective Hamiltonians with specific desired properties has become known as Floquet engineering. \cite{Rudner_2020, oka_floquet_2019}
Along these lines several ways to control matter with light have been proposed, for example, the manipulation of topologically non-trivial states, \cite{OkaAoki, Lindner2011, Kitagawa2011, wang_observation_2013, mciver_light-induced_2020, Decker_2019, sentef_theory_2015, Hubener2017, Fleckenstein2020} strongly correlated materials \cite{Bukov2016, Claassen2017, Kennes2018-FloquetChains, Mentink2015, Walldorf2019} and superconductors. \cite{Sentef2016, Knap2016, Kennes2017, Murakami2017,Porta2019, Kennes_2018} However, a fundamental problem for driving materials with classical light is heating, \cite{Murakami2017, DAlessio2014, Lazarides2014} which in many realistic setups prohibits versatile control.

To circumvent detrimental heating, control of materials through quantum light has recently been proposed. \cite{Kibis2011, wang_cavity_2019, Huebener2021, Ruggenthaler2018, FriskKockum2019}
The basic idea is to place a material into an optical cavity by which the light-matter coupling can be enhanced \cite{Dutra2004, FriskKockum2019} since the coupling is inversely proportional to the square-root of the effective mode volume \cite{Dutra2004, Li2020-Quantization}. One  can therefore bolster the coupling by manufacturing smaller devices, or by employing near-field enhancement effects.\cite{Maissen2014}
Through this enhancement of the coupling, vacuum fluctuations or few photon states of the cavity can already have a sizeable effect on the matter degrees of freedom, alleviating the need of strong classical driving fields.
In the emerging field of cavity engineering, ultra-strongly coupled light-matter systems have been realized based on different implementation schemes, starting from the first results obtained with microwave and optical cavities. \cite{Meschede1985, Thompson1992}
More recently, sizeable light-matter coupling (LMC) has been implemented in superconducting circuits,\cite{GU20171} and it is nowadays possible to couple few electrons to EM fields in split-ring resonators. \cite{Scalari2012, Keller2017, BallariniDeLiberato_2019}
These technological advances have led to the observation of LMC-controlled phenomena such as transport properties being tuned by polaritonic excitations \cite{ParaviciniBagliani2018} and
Bose-Einstein condensation of exciton-polaritons. 
\cite{kasprzak_bose-einstein_2006, Keeling_2020, byrnes_exciton-polariton_2014}
Another route to control matter by quantum light is to influence chemical reactions \cite{Anoop2016, Schaefer2021} through the selective enhancement of desired reactive paths and blocking of others.
In addition, there have been several proposals to influence superconductivity in a cavity, either by coupling cavity modes to the phonons involved in electronic pairing, \cite{sentef_cavity_2018} to magnons that are believed to form the pairing glue in cuprates, \cite{curtis2021} or by directly coupling to the electronic degrees of freedom. \cite{schlawin_cavity-mediated_2019, chakraborty_non-bcs-type_2020, Gao_Schlawin_2020, curtis_cavity_2019, Allocca2019} 
Concurrently, experimental evidence of cavity-enhanced superconductivity was recently reported, whose origin and interpretation are still under debate. \cite{thomas_exploring_2019}

To turn the question around and to add another facet to the problem of LMC, one can inversely ask: How can one engineer the light field of a cavity using matter? One prominent and widely discussed route is the realization of a superradiant phase in thermal equilibrium. \cite{Nataf2010, mazza_superradiant_2019, andolina_cavity_2019, ashida_demler, Schuler2020, Bernardis2018, Guerci2020, Reitz2021, stokes_uniqueness_2020}
Generally, systems that require a quantum-mechanical treatment of both light and matter will host hybrid states that mix light and matter degrees of freedom. \cite{Genet2021}
Describing such light-matter systems is a formidable challenge and often relies on using few-body simplifications.
For instance, describing matter through effective few-level systems has led to paradigmatic models such as the Dicke, Rabi or Jaynes-Cummings models. These simplified models capture certain aspects of the underlying physics well.\cite{Dicke1954, kirton_introduction_2019, Fox, Dutra2004,  frisk_kockum_ultrastrong_2019}
However, in order to capture collective phenomena of solid-state systems, a many-body description of the material is needed.
Efforts in this direction include first-principles approaches, such as the density functional reformulation of QED, \cite{Tokatly, Ruggy_2014, Pellegrini2015} generalized coupled cluster theory \cite{Haugland2020} or hybrid-orbital approaches. \cite{Buchholz2020, Nielsen2018} In addition, a recent work presents the analytic solution of the free 2D electron gas coupled to a cavity. \cite{Rokaj2020}

In this work, we introduce and study an exactly solvable quantum lattice model for a solid coupled to the quantized light field of a cavity.
 At the same time, we aim at connecting quantum-photon phenomena to previous results of Floquet engineering by investigating the quantum-to-classical crossover.
To this end, we focus on a tight-binding chain coupled to a single mode modelling a resonance of a cavity, through a quantized version of the Peierls substitution that was recently introduced.\cite{Li2020, sentef_quantum_2020, Dmytruk2021, kiffner_manipulating_2019} As we aim to describe solid-state systems, we are mainly interested in the thermodynamic (TD) limit of this model, but we also connect to prior finite system size studies. 
First, we determine the groundstate (GS) of the system.
By exact numerical means, we exclude the existence of an equilibrium superradiant phase, consistent with existing no-go theorems. \cite{Nataf2010, andolina_cavity_2019} We show explicitly that gauge invariance must be taken into account carefully to prohibit false signatures of a superradiant phase upon expanding the Peierls substitution in orders of the LMC. We then concentrate on the thermodynamic limit where the electronic groundstate is found to remain the Fermi sea of the uncoupled system centered at quasi-momentum $k = 0$ consistent with the findings of Rokaj et al.\cite{Rokaj2020}
Using this insight, we analytically determine the photonic GS of the system to be a squeezed state.
Additionally, an analytical expression for the electronic spectral function is given. With this we establish the quantum analogues to paradigmatic Floquet results, such as dynamical localization or the emergence of replica bands, and pinpoint the differences between the classical and quantum cases.
To make the connection to Floquet results explicit, we analyze the quantum-to-classical crossover and show that the nonequilibrium spectral function of the system approaches that of a classically driven system in the limit of strong driving. 
Finally, the current response to a spatially uniform external field, i.e., the optical conductivity, is calculated and a f-sum rule for cavity-coupled systems is identified.
The presence of the single cavity mode induces a non-complete suppression of the Drude peak that remains even in the TD limit. 
This result is consistent with that previously found by Rokaj et al.\cite{Rokaj2020} for the 2D electron gas.
We attribute this feature to the effective electron-electron interaction mediated by the cavity.
\section{Results}
\subsection{Model}\label{sec:model}
\begin{figure}[t]
    \includegraphics[scale = 1.]{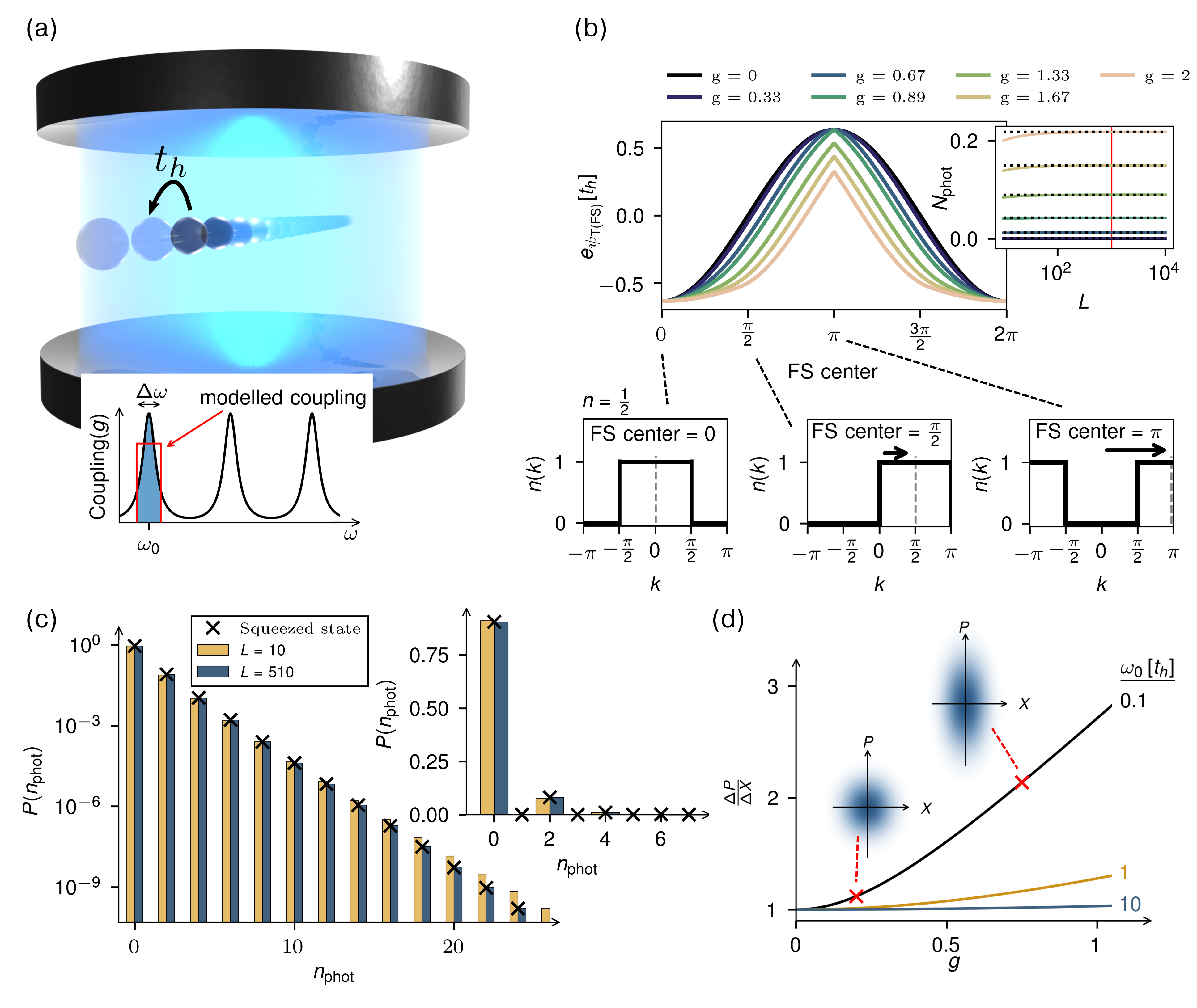}
    \caption{\textbf{Model and groundstate}.  \textbf{(a)}
    Illustration of the studied model: A one dimensional tight-binding chain with nearest neighbour hopping $t_h$ is coupled to the first transmittance resonance (blue shaded area) of a cavity at $\omega_0$.
    We model the frequency ($\omega$) dependent coupling (black line) as a box function (red line) and assume that its width $\Delta \omega \ll \omega_0$ to arrive at an effective single mode that couples strongly to the electrons (see the Model subsection under Results).
    \textbf{(b)} Energy density $e_{\psi_{\mathrm{T(FS)}}}$ according to Eq.~(\ref{eq:e}) (colored lines), with the electronic part of the wavefunction $\ket{\psi}_{f}$ chosen as  a single connected quasi-momentum region being occupied (Fermi sea, FS). 
    The minimum at wave-vector $k = 0$ coincides with that of the variational scheme described in the main text (see the Groundstate subsection under results and the Methods section) where we have used trial wave-functions with arbitrary distributions in momentum-space, i.e., not limited to a connected region.
    Inset: Average photon number $N_{\text{phot}} := \braket{a^{\dag}a}$ (colored lines) for varying coupling strength $g$, as function of the system size $L$.
    For all $g$ values shown, the number of bosons in the cavity converges at large $L$ to a finite value (black dashed lines).
    The red vertical line corresponds to the system size used in the main plot ($L = 1010$).
    $N_{\rm max}^{\rm boson} = 100$ has been used for the bosonic Hilbert space.
    \textbf{(c)} The exact probability distribution $P(n_{\rm phot})$ in logarithmic scale of the photon number is compared to the one given by a squeezed state (black crosses) for the groundstate of a chain of length $L = 510$ (blue bars) and $L = 10$ (yellow bars). Here the coupling constant is set to $g = 2 $ and $N_{\rm max}^{\rm boson} = 100$. 
    In the inset, the same quantity is plotted on a linear scale.
    \textbf{(d)} Ratio of variance of canonical momentum and coordinate operator $\Delta P/\Delta X$ (colored lines) as function of the coupling $g$ for three different values of $\omega_{0}$ and two representative squeezing ellipses for $g=0.2$ and $g=0.75$, respectively.
    }
    \label{fig:1}
\end{figure}

We consider a non-interacting tight-binding chain with nearest-neighbour hopping, as illustrated in Fig.~\ref{fig:1}(a).
The chain is coupled to the first transmittance resonance of a cavity.
We take into account a continuum of modes in the cavity but neglect modes that have a wave-vector with non-zero component in the direction of the chain as their coupling with the matter degrees of freedom will be strongly suppressed by the presence of the cavity.
This essentially amounts to the dipole approximation.
The frequency of the modes is confined to a small region of width $\Delta \omega$ around the resonance of the empty cavity at $\omega_0$ ($\Delta \omega \ll \omega_0$).
We therefore model these modes as all having the same frequency $\omega_0$.
Additionally, we assume that they couple to the chain with equal strength essentially replacing the frequency dependent profile of the coupling by a box function of width $\Delta \omega$ centered at $\omega_0$ (see Fig.~\ref{fig:1}(a)).
In Supplementary Note 1, we show that having selected $N$ modes, this setup results in one single mode strongly coupling to the electrons and $N-1$ uncoupled modes.
Hence, we model the system as electrons coupled to an effective single cavity mode that is spatially constant along the chain.
The corresponding Hamiltonian reads\cite{Li2020}
\begin{equation}
    H = \omega_0 \left(\ada + \frac{1}{2}\right) - \sum_{j = 1}^{L} \left[ t_{h} e^{-i \, \frac{g}{\sqrt{L}}(\aplusa)} \, c_{j + 1}^{\dagger} c_{j} + \text{h.c.} \right].
    \label{model}
\end{equation}
Here $c_{j}(c_{j}^{\dagger})$ is the fermionic annihilation (creation) operator at lattice cite $j$, and $a(a^{\dagger})$ is the bosonic annihilation (creation) operator of the single effective cavity mode. The latter are related to the quantized electromagnetic vector potential via $A=\frac{g}{\sqrt{L}}(a+a^{\dag})$,  with the convention $e=\hbar =c=1$ and $L$ the number of lattice sites.
We use periodic boundary conditions and set the lattice constant to $1$.
One can show that, within a few-band truncation, inclusion of the relevant effects of the LMC as well as gauge invariance are guaranteed by the quantized form of the Peierls substitution employed to set up the Hamiltonian given in Eq.~(\ref{model}).\cite{Li2020, sentef_quantum_2020, Dmytruk2021}
The coupling constant $g$ depends on the specifics of the system, such as the geometry and material composition of the cavity.
We keep the explicit dependence $1/\sqrt{L}$, instead of including it in the dimensionless coupling parameter $g$, in order to simplify the analysis of the thermodynamic limit.
In quasi-momentum space, the model takes the form

\begin{equation}
    H=\cos\left(\frac{g}{\sqrt{L}}(a+a^{\dag})\right) \mathcal{T}+\sin\left(\frac{g}{\sqrt{L}}(a+a^{\dag})\right) \mathcal{J}  + \omega_{0} \left(a^{\dagger}a + \frac{1}{2}\right),
     \label{eq:HKspace}
\end{equation}
where we have introduced the kinetic energy and current operators
\begin{equation}
    \begin{aligned}
    \mathcal{T} & := \sum_k - 2 t_{h} \cos(k) \, \cdc =: \sum_k \varepsilon_k \, \cdc\\
    \mathcal{J} & := \sum_k 2 t_{h} \sin(k) \, \cdc =: \sum_k v_k \, \cdc ,
    \end{aligned}
\end{equation}
and $\varepsilon_k$, $v_k$ are the band dispersion and band velocity at quasi-momentum $k$, respectively.
$c_k^{(\dag)}$ annihilates (creates) and electron at quasi-momentum $k$.
These expressions highlight the extensive number of constants of motion of the model, namely $\rho_k = c_k^{\dagger} c_k$ with 
$[\rho_k, H] = 0$ for all $k \in \text{BZ}$ (Brillouin Zone), which is a consequence of the spatially constant vector potential not breaking the lattice periodicity and preserving fermionic quasi-momentum in any electron-photon scattering process. \cite{Rokaj2020}
As a consequence, the eigenstates of the Hamiltonian can be factorized as
\begin{equation}
    H | \Psi \rangle = E_{\Psi}|\Psi \rangle \hspace{2mm}; \hspace{3mm}%
    |\Psi \rangle = \ket{\phi}_b\otimes\ket{\psi}_{f},
    \label{eq:eigenstatesFactorize}
\end{equation}
%
where
 $\ket{\phi}_b$ is the photonic part of the wavefunction, and $\ket{\psi}_f$ is an eigenstate of the electronic density operator $\rho = \frac{1}{L} \sum_k \cdc$.

\subsection{Groundstate}
\label{sec:GS}
We determine the GS of the system $\ket{\Psi_{\mathrm{GS}}}=\ket{\phi_{\mathrm{GS}}}_b\otimes\ket{\psi_{\mathrm{GS}}}_{f}$ in two different ways: (i) by a variational scheme that exploits the extensive number of constants of motion varying the electronic occupation and using exact diagonalization for the remaining non-harmonic bosonic system (see the Methods section) and (ii) by full exact diagonalization of the combined electronic and bosonic system (ED). 
The variational scheme can be performed for hundreds of lattice sites while
the ED calculations serve to verify the variational results for small system sizes.
Both numerical methods are exact in the sense that their accuracy is only limited by the cutoff of the maximum boson number in the Fock space $N_{\rm max}^{\rm boson}$.
This can, however, be chosen large enough to converge all calculations to arbitrary precision, making the results obtained with ED identical to those obtained with the variational method in the case of small system sizes.
Since the data reported in the plots has been acquired for system sizes too large for ED to handle, all reported results have been obtained with the variational scheme.

We consider a half-filled electronic system with
 $   n:=\langle\rho\rangle = \frac12,$ 
 and choose the cavity frequency $\omega_0 = t_h$, unless explicitly denoted otherwise.
Within  the variational scheme, we find that the electronic part of the GS wavefunction $|\psi_{\mathrm{GS}} \rangle_f$ is the Fermi sea (FS) around $k = 0$ even at non-zero $g$.
In Fig.~\ref{fig:1}(b) we illustrate this for a subset of possible electronic configurations.
Here, following the procedure explained in the Methods section, we take as fermionic trial wavefunctions $|\psi_{\mathrm{T(FS)}}\rangle_f$ only connected regions in $k$-space centered at different positions (FS center). Then we numerically determine the GS energy $E_{\psi_{\mathrm{T(FS)}}}$ of the resulting bosonic hamiltonian $H_{\psi_{\mathrm{T(FS)}}} = \,_f\langle \psi_{\mathrm{T(FS)}} | H |\psi_{\mathrm{T(FS)}} \rangle_f$. In Fig.~\ref{fig:1}(b) we show the energy density
\begin{equation}\label{eq:e}
    e_{\psi_{\mathrm{T(FS)}}} = \frac{E_{\psi_{\mathrm{T(FS)}}}}{L}
\end{equation}
as a function of the center of the connected region (FS center).
The energetic minimum always remains at the FS centered around $k = 0$ for all considered coupling values.
This shows that the fermionic part of the GS wavefunction remains unchanged upon turning on a coupling to the bosonic mode, a result that is consistent with the two-dimensional electron gas considered by Rokaj et al.\cite{Rokaj2020}
The unbiased variational scheme (see the Methods section) is not limited to connected regions in $k$-space, and a full variation in electronic state space confirms the unshifted Fermi sea as the true ground state. 

We now discuss the bosonic part of the wavefunction, $|\phi_{\mathrm{GS}} \rangle_b$. 
To this end, we define the photon number eigenstates as $a^{\dag}a \ket{n_{\rm phot}} = n_{\rm phot}\ket{n_{\rm phot}}$ and introduce the probability distribution $P(n_{\rm phot}) := |\braket{n_{\rm phot}|\phi_{GS}}|^{2}$ of finding $n_{\rm phot}$ photons in the GS.

$P(n_{\rm phot})$ for $g = 2$ (Fig.~\ref{fig:1}(c)) shows that only even number states contribute, implying that the bosonic wavefunction has a probability distribution that is incompatible with a coherent state.
Instead, $P(n_{\rm phot})$ agrees perfectly with a squeezed state with the same average photon number, indicated by the black crosses in Fig.~\ref{fig:1}(c).
This finding does not change qualitatively for different values of $g$. 
In the inset of Fig.~\ref{fig:1}(b) we show the scaling of the average photon number in the GS, $N_{\text{phot}} = \langle \ada \rangle$.
$N_{\text{phot}}$ is found not to grow extensively with the system size, which excludes the existence of a superradiant phase.

Put differently, the absence of a superradiant phase implies that the expectation value of the bosonic operators in the GS does not scale with the system size.
This allows us to perform a scaling analysis of contributions to the GS energy
\begin{equation}
\begin{aligned}
    \langle \Psi_{\mathrm{GS}} | H | \Psi_{\mathrm{GS}} \rangle &= \underbrace{\langle \Psi_{\mathrm{GS}} | \omega_0 \left(a^{\dag}a + \frac{1}{2}\right) | \Psi_{\mathrm{GS}} \rangle}_{\sim 1} + %
\underbrace{\langle \Psi_{\mathrm{GS}} | \mathcal{T} | \Psi_{\mathrm{GS}} \rangle}_{\sim L} + %
\underbrace{\langle \Psi_{\mathrm{GS}} | \frac{g}{\sqrt{L}} \left(a^{\dagger} + a\right) \mathcal{J} | \Psi_{\mathrm{GS}} \rangle}_{\sim \sqrt{L}}\\
&-\underbrace{\langle \Psi_{\mathrm{GS}} | \frac{1}{2} \frac{g^2}{L} \left(a^{\dagger} + a\right)^2 \mathcal{T} | \Psi_{\mathrm{GS}} \rangle}_{\sim 1}%
+ \mathcal{O}\left(\frac{1}{\sqrt{L}}\right).
\end{aligned}
\label{eq:scalingAnalysis}
\end{equation}
In the TD limit, the GS energy is entirely composed of terms that are at most quadratic in the photon field amplitude $A = \frac{g}{\sqrt{L}} (\aplusa)$. In order to simplify the following discussion, we diagonalize the Hamiltonian up to quadratic ($A^2$) order by a combined squeezing and displacement transformation yielding (see Supplementary Note 2)
\begin{equation}\label{eq:Hamiltonian_diag}
    H^{\text{D}} = \mathcal{W}[\mathcal{T}] \left( \beta^{\dag}\beta + \frac{1}{2}\right) + \mathcal{T} - \frac{g^2 \omega_0  \mathcal{W}[\mathcal{T}]^{-2}}{L}\mathcal{J}^2%
    \hspace{2mm}; \hspace{3mm} \mathcal{W}[\mathcal{T}] = \omega_0 \sqrt{1 - 2 \frac{g^2}{L\omega_0} \mathcal{T}}.
\end{equation}
where $\beta^{(\dag)}$ annihilates (creates) a coherent squeezed state. \cite{Kennes2017}
In terms of the original creation and annihilation operators of the unsqueezed cavity photons, the corresponding squeezed-state operators are given as
\begin{equation}
    \begin{aligned}
    \beta^{\dagger} &= \cosh \left(\frac{1}{2} \ln\left( \frac{\mathcal{W}[\mathcal{T}]}{\omega_0} \right) \right) \left( a^{\dagger} + \frac{g \, \omega_0 \mathcal{W}[\mathcal{T}]^{-2}}{L} \mathcal{J} \right)%
    + \sinh \left(\frac{1}{2} \ln\left( \frac{\mathcal{W}[\mathcal{T}]}{\omega_0} \right)\right) \left( a + \frac{g \, \omega_0 \mathcal{W}[\mathcal{T}]^{-2}}{L} \mathcal{J} \right),\\[8pt]
    \beta &= \cosh \left(\frac{1}{2} \ln\left( \frac{\mathcal{W}[\mathcal{T}]}{\omega_0} \right) \right) \left( a + \frac{g \, \omega_0 \mathcal{W}[\mathcal{T}]^{-2}}{L} \mathcal{J} \right)%
    + \sinh \left(\frac{1}{2} \ln\left( \frac{\mathcal{W}[\mathcal{T}]}{\omega_0} \right)\right) \left( a^{\dagger} + \frac{g \, \omega_0 \mathcal{W}[\mathcal{T}]^{-2}}{L} \mathcal{J} \right).
    \end{aligned}
\end{equation}
The last term in $
    H^{\text{D}} $ of Eq.~\eqref{eq:Hamiltonian_diag} highlights that the cavity induces an effective electron-electron interaction.

Knowing that the electronic part of the GS wavefunction is the unshifted FS, we define the expectation value of the electronic kinetic energy density and current density in the GS as
\begin{equation}
    \begin{aligned}
            t_{\mathrm{GS}} &= \frac{\,_f\langle\psi_{\mathrm{GS}} | \mathcal{T} | \psi_{\mathrm{GS}}\rangle_f}{L} < 0,\\
        j_{\mathrm{GS}} &= \frac{\,_f\langle\psi_{\mathrm{GS}} | \mathcal{J} | \psi_{\mathrm{GS}}\rangle_f}{L} = 0,
    \end{aligned}
    \label{eq:TGSJGS}
\end{equation}
and the dressed cavity frequency as
\begin{equation}
    \tilde{\omega} = \mathcal{W}[t_{\mathrm{GS}}] = \omega_0 \sqrt{1 + 2 \frac{g^2}{\omega_0 } |t_{\mathrm{GS}}|}.
\end{equation}
The bosonic part of the GS wavefunction is then given by the GS of the electronically renormalized bosonic Hamiltonian
\begin{equation}
    H^{\mathrm{D}}_{b} = \,_f\langle \psi_{\mathrm{GS}}| H^{\mathrm{D}} | \psi_{\mathrm{GS}} \rangle_f = \tilde{\omega} \left( \beta^{\dag}\beta + \frac{1}{2} \right) - |t_{\mathrm{GS}}| L
\end{equation}
which is a squeezed vacuum state $|\phi_{\mathrm{GS}}\rangle_b$\cite{Fox, bagchi_pedestrian_2020, Rabl2004, Glauber1991}
that is connected to the bare cavity vacuum $|0\rangle$ through a squeezing transformation,
\begin{equation}
    |\phi_{\mathrm{GS}}\rangle_b = e^{\frac{1}{2}  \left(\zeta^{*} a^2 - \zeta(a^{\dagger})^2 \right)}|0\rangle.
\end{equation}
The squeeze factor $\zeta$\cite{Wall_optics} is given by (see Supplementary Note 2) 
\begin{equation}
    \zeta = \frac{1}{2} \ln \left(\frac{\tilde{\omega}}{\omega_0}\right).
    \label{eq:squeeze}
\end{equation}

The squeezed state that was  numerically observed to match the exact $P(n_{\mathrm{phot}})$ for the GS Fig.~\ref{fig:1}(c) corresponds precisely to the squeeze factor $\zeta$ defined in Eq.~(\ref{eq:squeeze}). 
In Fig.~\ref{fig:1}(d) we show how the amount of squeezing depends on the cavity coupling strength $g$. Defining $X:= \left(\aplusa\right)$ and $P := i\left(a^{\dag}-a\right)$, and $\Delta \mathcal{O}= \sqrt{\braket{\mathcal{O}^{2}}- \braket{\mathcal{O}}^{2}}$ for a generic operator $\mathcal{O}$, a squeezed state minimizes the Heisenberg uncertainty $\Delta P \Delta X = 1$. The ratio
\begin{equation}
    \frac{\Delta P}{\Delta X}=e^{2\zeta}=\frac{\tilde{\omega}}{\omega_{0}}
\end{equation}
characterizes the degree of squeezing. \cite{Wall_optics, Fox} 
The squeezing of the vacuum is reminiscent of the finding by Ciuti et al.,\cite{Ciuti2005} which was obtained for a different light-matter model.
It has recently become possible to directly measure the vacuum fluctuations inside a cavity, \cite{Riek420, Benea-Chelmus2019} which enables experimental tests of our prediction.

\subsection{False superradiant phase transition in the approximate model}

\begin{figure}[t!]
    \includegraphics[scale = 1.]{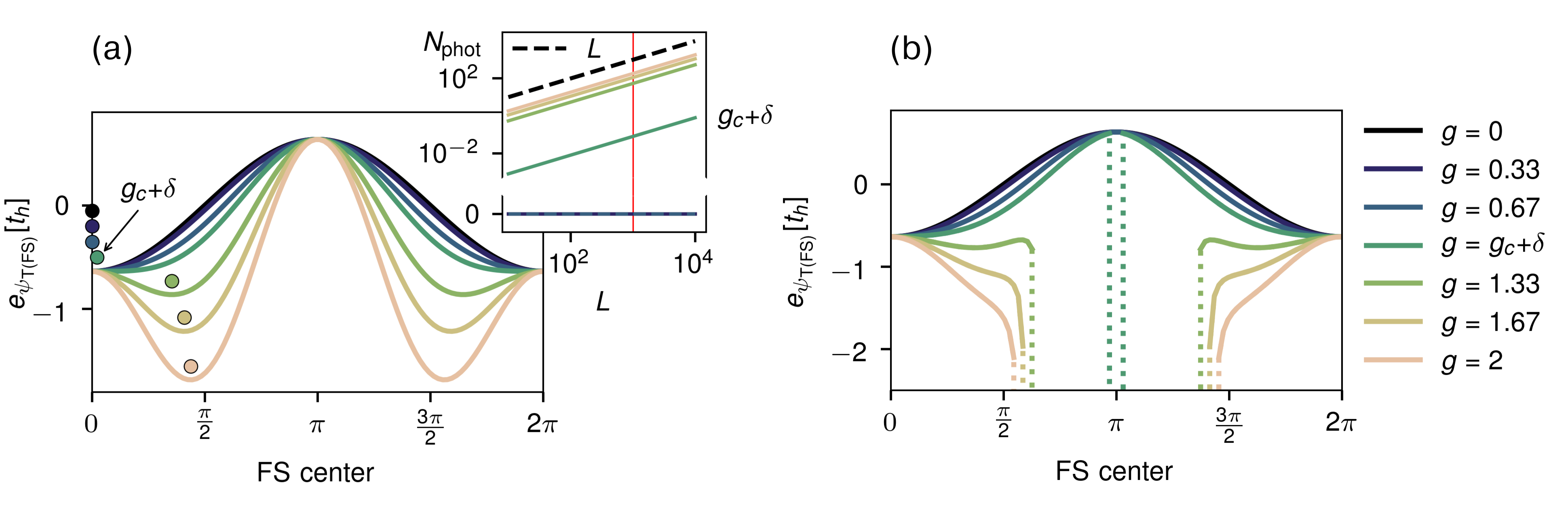}
    \caption{\textbf{False superradiance and instability for the truncated Hamiltonian.} \textbf{(a)} Minimum energy density $e_{\psi_{\mathrm{T(FS)}}}$ Eq.~(\ref{eq:e}) (colored lines) of the Hamiltonian truncated at first order for an electronic wavefunction being a single connected occupied region in $k$-space, as function of the shift of the Fermi sea (FS). The position of one minimum of the curves is indicated by a circle. At a critical coupling strength $g_c = \sqrt{\frac{\pi \omega_{0}}{4t_{h}}}$ the center of the Fermi sea realizing the minimal energy moves to a finite $k$-value which is illustrated by the small shift of the minimum of the curve corresponding to $g = g_c + \delta$ where $\delta = 0.001$. Inset: Average photon number $\braket{a^{\dag}a}$ (colored lines) for varying coupling strengths $g$ as function of the system size $L$. Above the critical value $g_{c}$, superradiant scaling of the photonic occupancy sets in. The vertical red line denotes the system size used in the main plot ($L = 1010$). \textbf{(b)} Minimum energy density of the second-order truncated Hamiltonian (colored lines) as function of the shift of the Fermi sea (FS). 
    When the shift is sufficiently large such that the kinetic energy of the electrons is positive, it is possible to obtain a spectrum of the electronically renormalized bosonic Hamiltonian that is not bounded from below anymore, rendering the system unstable.
    The instability is indicated by the dotted line. Here $L = 1010$.}
    \label{fig:2}
\end{figure}
Next, we analyze the effect of truncating the Hamiltonian at first and second order in $A = \frac{g}{\sqrt{L}}(\aplusa)$ on the GS at finite $L$
\begin{equation}
\begin{aligned}
H^{1^{\mathrm{st}}}  &= \omega_0 \left(a^{\dag}a + \frac{1}{2}\right) +  \mathcal{T} + \frac{g}{\sqrt{L}} \left(a^{\dagger} + a\right) \mathcal{J}\\
H^{2^{\mathrm{nd}}}  &= \omega_0 \left(a^{\dag}a + \frac{1}{2}\right) +  \mathcal{T} + \frac{g}{\sqrt{L}} \left(a^{\dagger} + a\right) \mathcal{J} %
- \frac{1}{2} \frac{g^2}{L} \left(a^{\dagger} + a\right)^2 \mathcal{T}.
\end{aligned}
\label{eq:Happrox}
\end{equation}
For the first-order truncated Hamiltonian $H^{1^{\mathrm{st}}}$ we again determine the GS by the unbiased variational scheme (see Methods section).
The GS is given by a connected region in $k$-space that is, however, not always centered at $k = 0$.
This is shown in Fig.~\ref{fig:2}(a), where the energy density $e_{\psi_{\mathrm{T(FS)}}}$ (Eq.~(\ref{eq:e})) for $H^{1^{\mathrm{st}}}$ is evaluated as function of the FS shift, in analogy to our analysis in Groundstate subsection under Results.
Here both the energy density and the photon occupation are calculated analytically.
We find that at a critical coupling strength $g_c$ there is a phase transition to a GS hosting a finite current signified by the shift of the FS, Fig.~\ref{fig:2}(a). 
This is complemented by an occupation of the cavity mode that scales linearly with $L$ as shown in the inset of Fig.~\ref{fig:2}(a)
as well as a non-zero expectation value in the TD limit of the field $\langle A \rangle = \frac{g \sqrt{L}}{ \omega_0} j_{\mathrm{GS}}$.
The critical coupling is given by $g_{c} = \sqrt{\frac{\pi \omega_0}{4 t_h}}$.
A symmetric or anti-symmetric combination of the degenerate GS wavefunctions (FS shifted either to the left or the right) would yield a net zero current restoring the inversion symmetry of the system but still result in a macroscopic occupation of the cavity mode. 
This transition is reminiscent of the one in the Dicke model, for which neglecting the diamagnetic ($A^2$) coupling yields a superradiant phase defined through $\langle A \rangle \sim \sqrt{N_{\rm emitter}}$ (where $N_{\rm emitter}$ is the number of emitters) yielding a macroscopically occupied photon mode, \cite{Dicke1954, kirton_superradiant_2018} which is absent for the full gauge-invariant coupling. \cite{Rzazewski1975}

In the lattice case, only the inclusion of coupling terms to all orders in $A$ of the the Peierls substitution guarantees gauge invariance.
If one instead includes only terms up to second order ($A^2$), a large coupling strength $g$ results in a spectrum of the Hamiltonian that is not bounded from below.
Fig.~\ref{fig:2}(b) is obtained in an analogous way to Fig.~\ref{fig:1}(b), but with energies calculated analytically, illustrating the absence of a GS above a critical coupling strength as follows: 
Fixing the electronic part of the wavefunction to be a shifted FS, an increased shift will yield a corresponding bosonic problem with a decreased frequency.
At some point the effective frequency vanishes, leading to the absence of a GS of the remaining bosonic problem beyond that point.
We indicate this point by a dotted line in Fig.~\ref{fig:2}(b).
This instability can be cured by including an arbitrarily small $A^4$ term, signalling the breakdown of the truncation.

States with a finite current, which have lower energy than the one with zero current when the energy is truncated after the first two orders of the LMC (see Fig.~\ref{fig:2}(b)), are moved to higher energies upon inclusion of all orders of the Peierls coupling (see Fig.~\ref{fig:1}(b)), which is a manifestation of gauge invariance.\cite{andolina_cavity_2019}
This explains the validity of our analytical results obtained including only the second order of the cavity field together with the electronic GS with zero current.
The instability discussed here, caused by truncation of the LMC after the second order, has previously been noted by Dmytruk and Schir\'o\cite{Dmytruk2021} in the context of a mean-field approach to a two orbital model.

\subsection{Momentum-resolved spectral function in the TD limit}
\label{sec:spectralFunction}
\begin{figure}
    \includegraphics[scale = 1.]{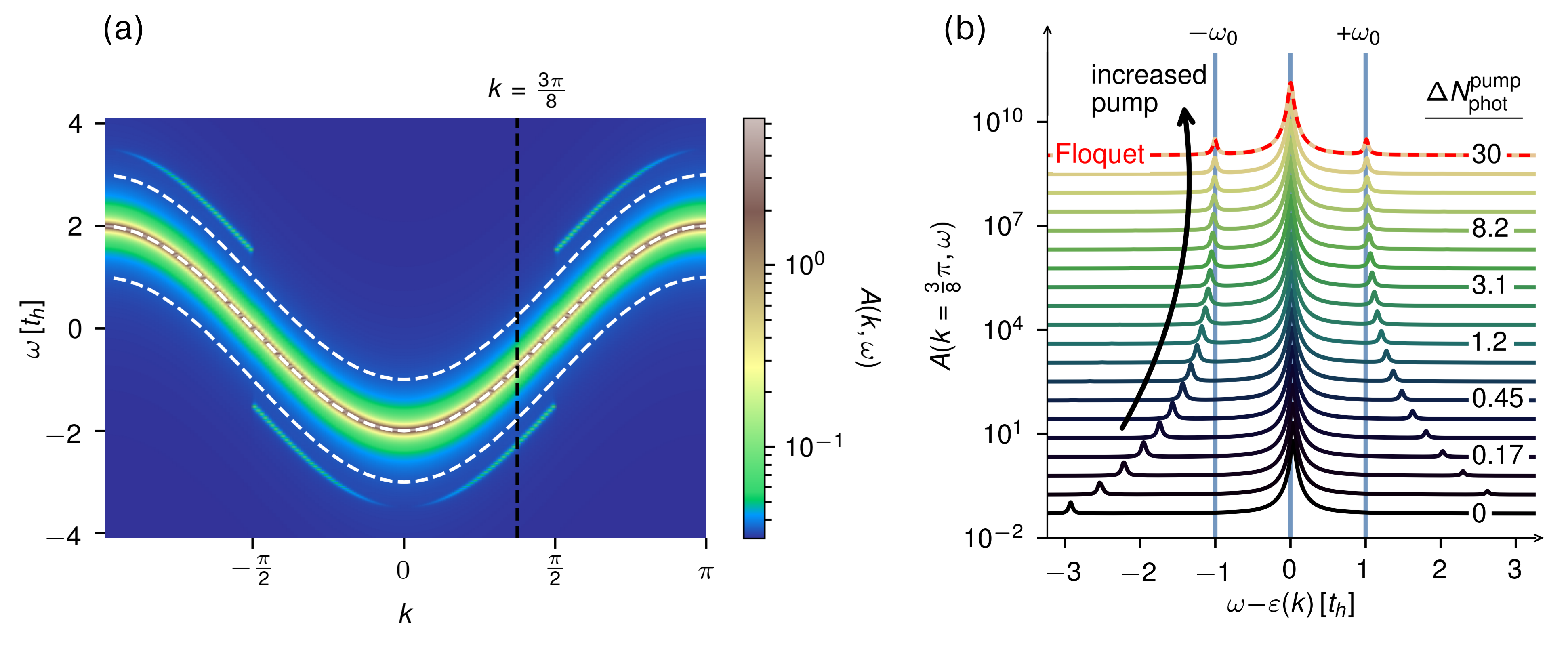}
    \caption{
    \textbf{Momentum-resolved spectral function in equilibrium and for a driven cavity.}
    \textbf{(a)} False-color plot of the momentum($k$)-resolved spectral function $A(k, \omega)$ Eq.~(\ref{eq:spectralFunction}) as function of frequency $\omega$ in units of the hopping amplitude $t_{\rm h}$ at $T = 0$.
    The central white dashed curve shows the bare electronic band.
    Replicas of the bare band offset by the bare cavity frequency $\pm \omega_0$ are shown by white dashed curves.
    The quantum replica bands seen in the false-color spectra are at an increased distance from the main band, which is set by the dressed cavity frequency $\tilde{\omega} > \omega_0$.
    The replica bands are below (above) the main band in the occupied (unoccupied) quasi-momentum regions, reflecting the overall particle-hole symmetry of the half-filled system.
    The dashed line at $k = \frac{3 \pi}{8}$ denotes the $k$-space position of the plot in panel (b). Here we consider $L = 170$, $g = 1$ and $N_{\rm max}^{\rm boson} = 50$, the delta functions of Eq.~(\ref{eq:spectralFunction}) are represented by Lorentzians  with broadening  $\eta = 0.025$.
    \textbf{(b)} Nonequilibrium time- and momentum-resolved spectral function according to Eq.(\ref{Non_Eq}) evaluated at $k=\frac{3\pi}{8}$ as a function of frequency ($\omega$) offset by the value of the dispersion $\varepsilon(k)$ at that $k$-point in units of the hopping amplitude $t_{\rm h}$ for several cavity pumping strengths, characterized by the displacement parameter $\alpha$ with $|\alpha|^2 = \Delta N_{\mathrm{phot}}^{\mathrm{pump}}$ (colored lines). $g^{2}\Delta N_{\mathrm{phot}}^{\mathrm{pump}}$ is kept constant, implying that $g\to0$ as the pumping $\Delta N_{\mathrm{phot}}^{\mathrm{pump}} \to \infty$. The black line corresponds to the ground state for $g = 2.5$ for which the y-axis reports the amplitude, while the coloured lines are vertically shifted for clarity 
    and follow the progressive occupation $\Delta N_{\mathrm{phot}}^{\mathrm{pump}}$ indicated on the right. 
    For increasing pump strength the side-bands become more symmetric and their position approaches  $\omega_0$ as $\tilde{\omega}\stackrel{g\to0}{\longrightarrow}\omega_0$. 
    For the largest pump $\Delta N_{\mathrm{phot}}^{\mathrm{pump}}$ the curve is overlaid with the Floquet result (red dashed line), that matches the pumped-cavity result.
    Here $L = 90$, $N_{\rm max}^{\rm boson} = 100$, and a Lorentzian broadening $\eta = 0.025$ has been included in the delta functions.
    }
    \label{fig:3}
\end{figure}
The effects of the cavity on electrons could be investigated via ARPES measurements.
For this reason, but also to pinpoint analogs to Floquet results, we calculate the electronic spectral-function defined as
\begin{equation}
    A(k, \omega) = - \frac{1}{\pi}\, \text{Im} \, G^{\text{R}}(k, \omega),
    \label{eq:defSpectralFunction}
\end{equation}
with
\begin{equation}
    G^{\text{R}}(k, \omega) = - \int_{0}^{\infty} dt \, i \langle \left[c_k(t), c_k^{\dag}\right]_+ \rangle e^{i \omega t}
    \label{eq:GR}
\end{equation}
where $[.]_+$ is the anti-commutator.
We evaluate the electronic part of the expectation value in Eq.~(\ref{eq:GR}) analytically by commuting the electronic creation and annihilation operators with the appearing time-evolution operators and replacing $\mathcal{T} \rightarrow t_{\mathrm{GS}}L$ and $\mathcal{J} \rightarrow j_{\mathrm{GS}}L = 0$ in the expression.
The remaining vector-matrix-vector product in the bosonic part of the Hilbert space is then evaluated numerically at each time $t$ and the result transformed to frequency space via a FFT.
The result is given in Fig.~\ref{fig:3}(a) for a chain of length $L=170$ including all orders of the Peierls coupling.

In the TD limit, we can use similar arguments to the ones previously utilized in the Groundstate subsection under Results to give an analytic expression for the electronic spectral function. 
No operator in the expectation value Eq.~(\ref{eq:GR}) creates a macroscopic number of photons. We can thus conclude by a similar scaling analysis as in Eq.(\ref{eq:scalingAnalysis}) that in the TD limit the time evolution can be written with the diagonal Hamiltonian Eq.~(\ref{eq:Hamiltonian_diag}).
The spectral function keeping leading $1/L$ corrections is analytically found to be
\begin{equation}
\begin{aligned}
    A(k, \omega) & = (1 - n_k) e^{- \frac{g^2 v_k^2 \omega_0}{L \tilde{\omega}^3}} \sum_{\ell} \frac{\left(\frac{g^2 v_k^2 \omega_0}{L \tilde{\omega}^3} \right)^{\ell}}{\ell !}
\delta\left(\omega - \varepsilon_k \left(1 - \frac{g^2}{2L}\frac{\omega_0}{\tilde{\omega}}\right) - \Sigma_k - \tilde{\omega} \ell\right)\\
& + \, n_k \, e^{- \frac{g^2 v_k^2 \omega_0}{L \tilde{\omega}^3}} \sum_{\ell} \frac{\left(\frac{g^2 v_k^2 \omega_0}{L \tilde{\omega}^3} \right)^{\ell}}{\ell !}
\delta\left(\omega - \varepsilon_k \left(1 - \frac{g^2}{2L}\frac{\omega_0}{\tilde{\omega}}\right) + \Sigma_k + \tilde{\omega} \ell\right).
\end{aligned}
\label{eq:spectralFunction}
\end{equation}
Here $n_k = \langle \rho_k \rangle$ and the self-energy $\Sigma_k$ is given by
\begin{equation}
    \Sigma_k = \frac{g^2 \omega_0}{\tilde{\omega}^2 L} v_k^2.
    \label{eq:defSigmaK}
\end{equation}
The details of the calculation are presented in Supplementary Note 3.
From Eq.~(\ref{eq:spectralFunction}) the spectral function of the unperturbed electrons,
\begin{equation}
    A(k, \omega) \overset{L \to \infty}{\to} A_0(k, \omega) = \delta(\omega - \varepsilon_k),
\end{equation}
is recovered in the limit $L \to \infty$.
From Eq.~(\ref{eq:Hamiltonian_diag}) one might expect a finite contribution to the electronic self-energy stemming from the coupling of a single electron to all other electrons collectively.
However, due to the form of the induced interaction, the single electron couples to the total current that vanishes identically in the GS.
Contributions to the spectral function beyond the described collective effect are small in the TD limit as highlighted in Eq.~(\ref{eq:spectralFunction}).
We discuss how this might be related to a short-coming of the single-mode approximation in the Discussion.

The spectral function Eq.~(\ref{eq:spectralFunction}) most prominently contains a sum over $\delta$ functions with distance $\tilde{\omega}$ between each other, given by the dressed instead of bare cavity frequency, which is a direct consequence of the quantum nature of the photons.
This is the quantum analog to the Floquet replica bands visible in Fig.~\ref{fig:3}(a).
Contrary to the Floquet replica bands, the quantum replica bands lie either above or below the main band, but only on one side for fixed quasi-momentum $k$ at zero temperature, depending on whether the respective momentum state is filled or empty.
This reflects the particle-hole symmetry of the half-filled system, in which a combined $\omega \rightarrow -\omega$ and $k \rightarrow k + \pi$ sublattice particle-hole transformation leaves the spectral function invariant.

Importantly, despite the fact that the cavity induces an effective all-to-all electron-electron interaction, there is no broadening of the $\delta$-peaks.
This is related to the vanishing momentum transfer of the interaction and the resulting fact that the Bloch states remain exact electronic eigenstates.
As a consequence, the interaction results in a purely real electronic self-energy $\Sigma_k$, leading to band renormalizations without broadening.

The presence of the cavity squeezes the band dispersion $\varepsilon_k$ by a factor $\left(1-\frac{g^2}{2L}\frac{\omega_0}{\tilde{\omega}}\right)<1$.
This is the quantum analog to the dynamical localization that leads to a suppression of the band width.
The band renormalization factor $1 - \frac{g^2 \omega_0}{2L \tilde{\omega}}$ is consistent to leading order in $\frac{1}{L}$ with the expectation value of the bosonic operator $\langle \cos \left( \frac{g}{\sqrt{L}} \left( \aplusa \right) \right) \rangle$ as a multiplicative factor to the kinetic energy of the electrons.
The electrons are thus effectively localized by coupling to the vacuum fluctuations of the electromagnetic field.

\subsection{Quantum to Floquet crossover}
\label{sec:QuantumToFloquet}

In the following, we analyze the quantum to classical crossover and recover known Floquet physics in the regime of $N_{\text{phot}} \to \infty$ and $g \to 0$, keeping $ g^2 \, \Delta N_{\mathrm{phot}}^{\mathrm{pump}} = \mathrm{const}$. 
The limit $g \to 0$ is needed in the crossover to lift the light-matter hybridization that would otherwise lead to the shifted frequency $\tilde{\omega}$ of an effective cavity mode which we identify as an intrinsic quantum effect.
The limit of strong pumping, keeping the coupling $g$ constant, is treated in Supplementary Note 4.

We employ a protocol where the cavity mode is coherently displaced with respect to the GS with displacement parameter $\alpha$
\begin{equation}
    |\alpha \rangle = e^{\alpha(a^{\dagger} - a)} | \phi_{\mathrm{GS}} \rangle_{b}.
\end{equation}
The photon number is thereby increased relative to the one in the GS by $|\alpha|^2 = \Delta N_{\mathrm{phot}}^{\mathrm{pump}}$. The coherent displacement considered here models the application of a laser pumping the cavity on time scales too short for the coupled system to follow. Thus, the laser is assumed to place the cavity into a squeezed coherent state in the limit of large system size.
The subsequent time evolution of the light-matter coupled system is considered from starting time $t = 0$. 
While for the equilibrium spectral function only the first two orders in $g$ of the Hamiltonian had to be taken into account, the time evolution is now affected by all orders of the Peierls coupling due to the occupation of the photonic mode that is macroscopic in the classical limit. 


We calculate the nonequilibrium spectral function, defined via the full double-time retarded Green's function,\cite{Freericks2009}
\begin{equation}\label{Non_Eq}
\begin{aligned}
    &A_{\text{non-eq.}}(k, \omega) = \\
    &\frac{1}{\pi} \text{Im} \frac{1}{\tilde{\tau}} \int_{\Delta T - \frac{\tilde{\tau}}{2}}^{\Delta T + \frac{\tilde{\tau}}{2}} \left[ \int_{0}^{\infty} i e^{i \omega_{0} (t - t')} \, _f\langle \psi_{\mathrm{GS}} | \otimes \langle \alpha | \left[c_k(t), c_k^{\dag}(t')\right]_+ | \alpha \rangle \otimes | \psi_{\mathrm{GS}} \rangle_f \, d\left(t - t'\right) \right] d\left(\frac{t + t'}{2}\right)
\end{aligned}
\end{equation}
where $\tilde{\tau} = \frac{2\pi}{\tilde{\omega}}$ is the period corresponding to the dressed cavity frequency.
The form is chosen in analogy to the diagonal elements of the Floquet representation of the GF.\cite{Tsuji2008}
Here we include a waiting time $\Delta T$ after the start of the real-time evolution, set to a large value with respect to the intrinsic timescale, $\Delta T = 200 \tilde{\tau}$, in the numerical simulation.
Otherwise the calculation is performed in the same manner as that for the equilibrium spectral function Eq.~(\ref{eq:GR}).
For comparison, we also consider the nonequilibrium spectral function of a classically driven system where the time evolution is governed by the Hamiltonian
\begin{equation}
    H^{\text{c}}(t) = -\sum_{j} t_{h} e^{- i  A(t) } c_{j + 1}^{\dagger} c_j + h.c. 
    \label{eq:HSemiClassical}
\end{equation}
In this case, we couple the chain to the classical field
 $   A(t) = A_0 \sin(\omega_0 t),$
that oscillates with the eigenfrequency of the unperturbed cavity $\omega_0$.
Similar to the quantum case, we calculate the nonequilibrium spectral function according to
\begin{equation}
\label{eq:classicalspec}
\begin{aligned}
    &A_{\text{Floquet}}(k, \omega) = \\
    &\frac{1}{\pi} \text{Im} \frac{1}{\tau} \int_{- \frac{\tau}{2}}^{\frac{\tau}{2}} \left[ \int_{0}^{\infty} i e^{i \omega (t - t')} \, _f\langle \psi_{\mathrm{GS}} | \left[c_k(t)_{H^{\text{c}}(t)}, c_k^{\dag}(t')_{H^{\text{c}}(t)} \right]_+ | \psi_{\mathrm{GS}} \rangle_f \, d\left(t - t'\right) \right] d\left(\frac{t + t'}{2}\right)
\end{aligned}
\end{equation}
where $\tau = \frac{2 \pi}{\omega_0}$.
Here $(.)(t)_{H^{\text{c}}}$ denotes the time dependence governed by the semi-classical Hamiltonian Eq.~(\ref{eq:HSemiClassical}).
The spectral function fulfills
\begin{equation}
    A_{\text{Floquet}}(k, \omega + m \omega_0)|_{\omega \in \left(-\frac{\omega_0}{2}, \frac{\omega_0}{2}\right]} = - \frac{1}{\pi} \text{Im} \, G_{mm}(\omega)
\end{equation}
with $G_{mm}(\omega)$ the diagonal part of the \textit{Floquet representation} of the GF.\cite{Tsuji2008}

We show the evolution from quantum to Floquet spectra for a representative quasi-momentum $k = \frac{3 \pi}{8}$ inside the FS in Fig.~\ref{fig:3}(b).
In the extreme quantum case (GS) the replica band only appears below the main band.
Furthermore, it is not located at the bare cavity frequency $\omega_0$ but at the eigenfrequency of the coupled light-matter system $\tilde{\omega}$. 
By contrast, as the classical limit is approached, the symmetry of the replica bands is restored and their position moves to $\omega_0$.
For the largest displacement ($\Delta N_{\mathrm{phot}}^{\mathrm{pump}} = 30$) the spectrum matches precisely the Floquet spectrum.
The fact that the system experiences no heating during the driving is a direct consequence of the absence of electron-electron interactions and the corresponding macroscopic number of constants of motion.
\subsection{Optical conductivity}
\label{sec:methodsConductivity}

\begin{figure}
    \includegraphics[scale = 1.]{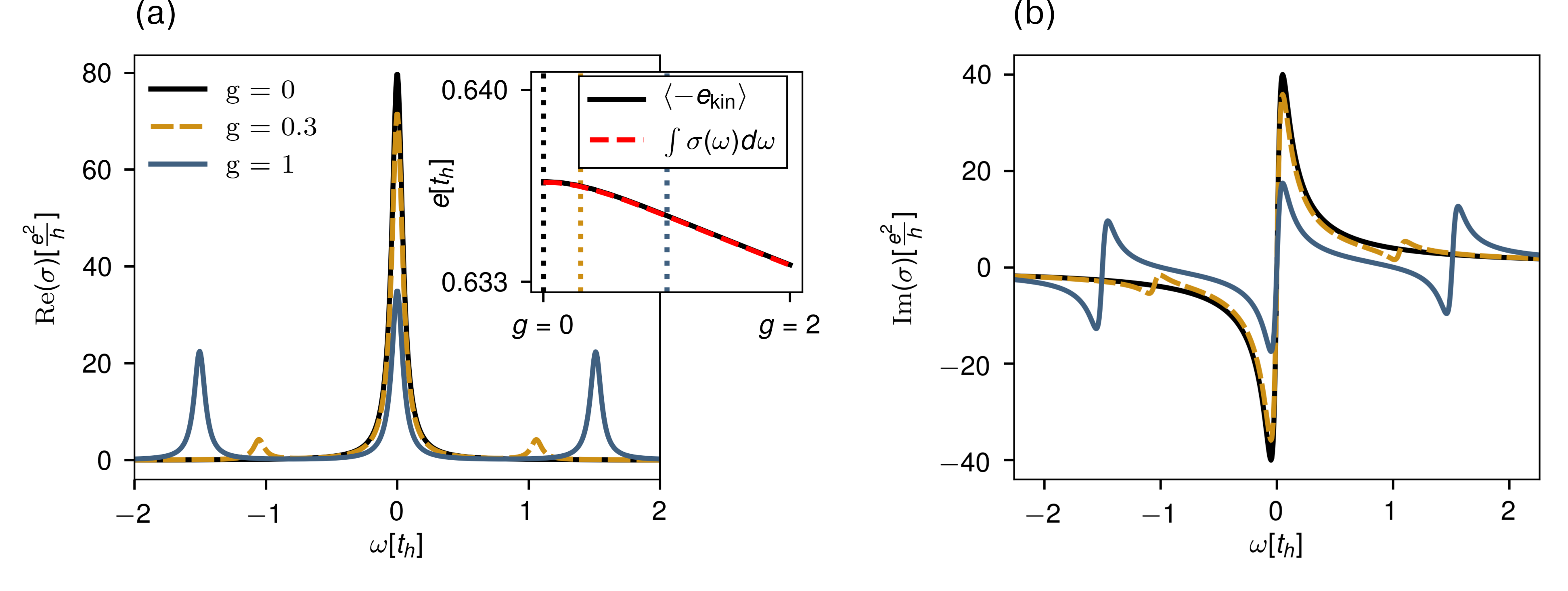}%
    \caption{
    \textbf{Optical conductivity}
    (\textbf{a}) Real part of the conductivity $\rm Re (\sigma)$, Eq.~(\ref{eq:Conductivity}) in units of half the conductance quantum $\frac{e^2}{h}$, for strong ($g=1$, dark blue line) and intermediate ($g=0.3$, dashed yellow line) couplings as a function of frequency $\omega$ in units of the hopping amplitude $t_{\rm h}$. The result for $g = 0$ is shown for comparison (black line).
    The Drude peak is suppressed with increasing $g$, and two side peaks appear at the same time.
    The inset shows the negative effective kinetic energy $\langle e_{\rm kin} \rangle$ (black line) and the integrated conductivity $\int \sigma(\omega) d\omega$ (red dashed line). The vertical dashed lines indicate the coupling strengths from the main plot. They match fulfilling the f-sum rule Eq.~(\ref{eq:sumrule}), here we set $L=170$, $N_{\rm max}^{\rm boson} = 50$ and a Lorentzian broadening $\eta=0.05$.
    \textbf{(b)} Corresponding imaginary parts of the conductivity $\rm Im (\sigma)$ (Eq.~(\ref{eq:imconductivity})).
    Again the central $\frac{1}{\omega}$ feature is suppressed and two side features appear at $\omega = \pm \tilde{\omega}$.
    }
    \label{fig:4}
\end{figure}

In order to discuss the impact of the light-matter coupling on a paradigmatic electronic two-particle response function, we compute the optical conductivity using the standard Kubo formalism.\cite{Rokaj2020, Amelio2021}
To this end the cavity-chain system is coupled to a spatially uniform external field $A_{\text{ext}}(t)$, in addition to the quantized cavity field.
The resulting optical conductivity in the long-wavelength limit is obtained in the standard form\cite{Scalapino_1993}
\begin{equation}
    \sigma(\omega) = - \frac{- \langle e_{\text{kin}} \rangle - \Lambda(q = 0, \omega)}{i\left(\omega + i0^+\right)},
\label{eq:conductivityGenerically}
\end{equation}
where
\begin{equation}
    e_{\text{kin}} = \frac{1}{L} \cos\left( \frac{g}{\sqrt{L}} \left(\aplusa\right) \right) \mathcal{T}
\end{equation}
is the effective kinetic energy density of the electrons in the cavity-modified GS, and
$\Lambda$ is the current-current correlator
\begin{equation}
    \Lambda(q = 0, \omega) = -\frac{i}{L} \int_{0}^{\infty} dt \,  \,e^{i \omega t} \langle \left[j_{q = 0}^{p}(t), j_{q = 0}^p\right] \rangle,
\end{equation}
with $j_{q = 0}^{p}$ the paramagnetic current density operator at $q = 0$. The latter is obtained from the charge continuity equation as
\begin{equation}
    j_{q = 0}^{p} = -\cos\left(\frac{g}{\sqrt{L}} (\aplusa)\right) \sum_k 2 t_{h} \, \sin (k) \cdc 
    - \sin \left(\frac{g}{\sqrt{L}} \left(\aplusa\right)\right) \sum_k \, 2 t_{h} \cos (k) \cdc.
\end{equation}
We evaluate Eq.~(\ref{eq:conductivityGenerically}) numerically for $L = 170$ and finite broadening $0^+ \rightarrow 0.05$. 
The result is shown in Fig.~\ref{fig:4}(a)-(b).

One can gain additional insight into the properties of the optical conductivity by evaluating it analytically in the TD limit.
For the real part of the conductivity we find
\begin{equation}
    \text{Re} \, {\sigma(\omega)} = D \delta(\omega) + \sigma_{\text{reg}}(\omega),
\label{eq:Conductivity}
\end{equation}
where the Drude weight $D$ is given as
\begin{equation}
    \frac{D}{\pi} = |t_{\text{GS}}| \left( 1 - \frac{g^2}{2 L} \frac{\omega_0}{\tilde{\omega}} - 2 \frac{g^2 \omega_0}{ \tilde{\omega}^2} |t_{\mathrm{GS}}| \right).
    \label{eq:drudeWeight}
\end{equation}
The second term in the brackets in Eq.~(\ref{eq:drudeWeight}) derives from the squeezing of the band, previously coined quantum dynamical localization, subsection Momentum-resolved spectral function in the TD limit under Results, and vanishes in the TD limit. 
The last term originates from the current-current correlator and remains finite even in the TD limit, resulting in a partial suppression of the  Drude weight.
In contrast to the spectral function considered in the subsection Momentum-resolved spectral function in the TD limit under Results, modifications to the optical conductivity remain finite even in the TD limit since the perturbation of the system within the linear response framework enables a contribution from the induced electron-electron interaction.
Writing
\begin{equation}
    \gamma = \frac{\omega_p^2}{\omega_0^2 + \omega_p^2} \hspace{1mm};\hspace{3mm} \omega_p^2 =  2 g^2 \omega_0|t_{\mathrm{GS}}|
\end{equation}
we find for $D$ in the TD limit
\begin{equation}
    D = D_0 (1 - \gamma) \hspace{1mm} ; \hspace{3mm} 0 \leq \gamma \leq 1
\end{equation}
where $D_0$ is the Drude weight of the uncoupled chain.
This is consistent with the findings Rokaj et al.\cite{Rokaj2020} for an electron gas.
For the second contribution $\sigma_{\mathrm{reg}}$ in Eq.~(\ref{eq:Conductivity}) one finds
\begin{equation}
    \frac{\sigma_{\text{reg}}(\omega)}{\pi} = \frac{g^2 \omega_0}{\tilde{\omega}^2} t_{\text{GS}}^2 %
    (\delta(\omega + \tilde{\omega}) + \delta(\omega - \tilde{\omega})).
\end{equation}
Two side-peaks at $\omega = \pm \tilde{\omega}$ appear that balance the suppression of the Dude weight.
These effects are illustrated in Fig.~\ref{fig:4}(a).

The inset of Fig.~\ref{fig:4}(a) shows that the real part of the conductivity satisfies the f-sum rule, similar to other electron-boson models\cite{Alvermann2010},
\begin{equation}
    \frac{D}{\pi} + \int_{-\infty}^{\infty} \sigma_{\text{reg}}(\omega) \, d\omega = - \langle e_{\text{kin}} \rangle,
    \label{eq:sumrule}
\end{equation}
which is also evident from the corresponding analytical expression. 

For completeness, we also state the imaginary part of the conductivity 
\begin{equation}\label{eq:imconductivity}
    \text{Im} \, {\sigma(\omega)} = t_{GS} \frac{1}{\omega} \left(1 - \frac{g^2}{2L} \frac{\omega_0}{\tilde{\omega}}\right) %
    + \frac{g^2 \omega_0}{\tilde{\omega}} t_{\text{GS}}^2 \frac{1}{\omega} \left( \frac{1}{\omega - \tilde{\omega}} - \frac{1}{\omega + \tilde{\omega}} \right).
\end{equation}
which fulfills the usual Kramers-Kronig relation $\text{Im} \, \sigma(\omega)=-\frac{1}{\pi}\mathcal{P}\int_{-\infty}^{\infty}\frac{\text{Re} \, \sigma(\omega ')}{\omega ' - \omega}d\omega '$ and is shown in Fig.~\ref{fig:4}(b).
Similar to the real part we find a suppression at $\omega = 0$ and shakeoff features at $\omega = \pm \tilde{\omega}$.

\section{Discussion}
\label{sec:discussion}
In this work, we have discussed a tight-binding chain coupled to a single spatially constant cavity mode.
The exact solution of this model is enabled by the macroscopic number of constants of motion that results from the absence of momentum transfer between photons and electrons in the long-wavelength limit.
Consequently, the GS of the system is a product state of electrons and photons (subsection Groundstate under Results).

Removing these constants of motion, either through relaxing the dipole approximation or including an electron-electron interaction, is expected to lead to interesting new results.
It is well known that a one-dimensional system with local interactions is susceptible to form a charge density wave at zero temperature. \cite{Giamarchi2003}
The effective interaction induced by the cavity considered in this work does not lead to such a symmetry-broken GS, since it is  featureless.
Including local interactions, it would therefore be interesting to study the effect of the cavity on charge-ordered phases.
An important consequence of the non-interacting limit is the absence of heating in the semi-classically driven case described in the subsection Quantum to Floquet crossover under Results.
In an interacting setup, a continuous classical drive would heat up the system eventually leading to an infinite temperature state.
On the other hand, an initial coherent state of the cavity will dissipate energy into the system leading to a decay of its amplitude.
For these reasons, the comparison made in the subsection Quantum to Floquet crossover under Results will only hold on time-scales much shorter than the time it takes for the system to heat up.
Previous works noted that even when including electron-electron interactions but neglecting any momentum transfer by the cavity photons, a factorized wave-function might still be suitable for a description of the system as the corresponding mean field picture becomes exact in the TD limit.\cite{Dmytruk2021, mazza_superradiant_2019, andolina_cavity_2019}

Relaxing the dipole approximation would lead to a finite-ranged but non-local effective electron-electron interaction, which opens new opportunities for inducing or modifying materials properties.\cite{Gao2021}
Through this, also existing no-go theorems related to superradiance would be circumvented, possibly making it worthwhile to revisit the question whether an equilibrium photon condensate can exist.\cite{andolina_cavity_2019, Dmytruk2021, katharine_collective_2020}

In order to describe realistic experimental situations, a continuum of modes needs to be included, where also the wave-vector in the direction of the chain is a continuous variable.
As a first approximation one might, as we did earlier for the orthogonal directions, treat these modes as identical.
For this case the principle of collective strong coupling that we describe in Supplementary Note 1 applies, leading to a mere renormalization of parameters.\cite{Rokaj2020}
However, macroscopically many modes coupled to all electrons at once will lead to unphysical effects like a diverging effective mode energy.
To remedy this also the dipole approximation would need to be relaxed making all but the zeroth mode couple to a microscopic quantity.

We have furthermore calculated the single-particle Green's function analytically (subsection Momentum-resolved spectral function in the TD limit under Results).
Here we found that in the limit $L \to \infty$ we recover the bare spectral function of the uncoupled electrons indicating that corrections due to the presence of the cavity vanish in the TD limit.
We pointed out that a possible mean-field term does not contribute due to the current in the GS having zero expectation value, $\langle \mathcal{J} \rangle = 0$.
Corrections beyond this are small in the TD limit which we attribute to the vanishing energy density of the single mode signified by $\frac{g}{\sqrt{L}} \to 0$ in that limit.
Supplementary Note 1 shows how such corrections could be reconciled through a collective coupling effect, reminiscent of previously discussed collective (vibrational) strong coupling,\cite{shalabney_coherent_2015, du_can_2021, sidler_perspective_2021, frisk_kockum_ultrastrong_2019, sidler_perspective_2021} when retaining many modes corresponding to a finite energy density in the TD limit which is reflected in the replacement $\frac{g}{\sqrt{L}} \to \frac{g \sqrt{N}}{\sqrt{L}}$.
This argument, however, requires further consideration such as the relaxation of the dipole approximation as mentioned above, to arrive at a mathematically rigorous conclusion. Such a calculation goes beyond the scope of this work.

The analytical expression for the single-particle Green's function derived in this work might provide the basis for future studies by building a many-body perturbation theory around this solution to investigate many-body instabilities diagrammatically, such as superconductivity. 
Note that the here considered system does not host polaritons since there are no collective bosonic excitations in our model such as plasmons, excitons or phonons as would be the case in a multi-band system. \cite{Dmytruk2021, katharine_collective_2020} Accordingly, no signatures of such quasi-particles show up in the electronic spectral function.
Using insights from the squeezing transformation, it might be possible to treat systems with two different bosonic modes analytically.
One interesting prospect is to include an optically active phonon into the model that couples quadratically to the electrons. \cite{Kennes2017, Sentef2016, Buzzi2020}
Extending the here-presented analytical methods to a bimodal squeezing, it might be possible to analytically obtain GS properties and signatures in electronic spectra of the coupled bosonic modes. This could open up a pathway to realize multi-mode squeezed states, with important applications to quantum information.\cite{RevModPhys.77.513}
In a similar spirit, one could also study two distinct photonic cavity modes and search for signatures of the matter-induced photon-photon interaction on the basis of the exactly solvable model put forward in the present work.

Concerning the connection to experiments, a temperature lower than the eigenfrequency of the cavity is needed in order for our zero-temperature calculations to hold qualitatively. For a resonance at $\omega_0 = 0.41 \mathrm{THz}$ as used in a recent cavity setup\cite{zhang_collective_2016} this would correspond to temperatures well below $3.1\mathrm{K}$.
The validity of the dipole approximation depends on the specific experimental setup. However, a sample that is much smaller that the size of the cavity is necessarily needed\cite{ruggenthaler_quantum-electrodynamical_2018} which would be fulfilled for a cavity size on the order of $1\rm mm$ corresponding to the above mentioned resonance at $\omega_0 = 0.41 \mathrm{THz}$ when at the same time considering an atomic wire with a length in the sub micrometer range.
The electronic spectra calculated here (Fig.~\ref{fig:3}(a)) should in principle be observable in ARPES measurements.
A quality factor that ensures a linewidth that is smaller than the cavity frequency is required to observe the side bands, which appears within experimental reach. \cite{zhang_collective_2016}
We attributed the vanishing of corrections to the spectral function in the TD limit to the vanishing energy-density of the single mode in that limit.
In an experimental setup one naturally has a continuum of modes with finite energy density possibly retaining these corrections.
For small enough in-plane wave-vectors of the photons one might expect qualitative effects, such as the asymmetry of the shake-off bands in the quantum limit, to remain present also in this case.
However, some further work definitely needs to be dedicated to this aspect in order to support this claim.
The experimental observation of asymmetric shake-off bands would complement the successful demonstration of classical Floquet replica bands.\cite{wang_observation_2013}

Another prediction of the present work is the squeezing of the vacuum fluctuations in the GS consistent with predictions for other models.\cite{Ciuti2005, Glauber1991}
Recently progress in probing the vacuum fluctuations of light \cite{Riek420, Benea-Chelmus2019} puts an experimental confirmation of our prediction within reach.

Finally, a suppression of the Drude peak (Fig.~\ref{fig:4}(a)) has already been observed experimentally. \cite{ParaviciniBagliani2018}
It has previously been explained by Rokaj et al. \cite{Rokaj2020} via an analogous result to the one presented by us but for an electron gas instead of a tight-binding chain.
It is an interesting question why the effective cavity mode with vanishing energy density can influence the macroscopically many electrons in this particular case.
From our point of view, the reason lies in the induced electron-electron interaction that does not vanish in the TD limit and is probed indirectly through the optical conductivity.

\section{Methods}

\subsection{Variational scheme}
\label{sec:methodsVariationalScheme}

Here, we describe the variational scheme that we use to determine the exact GS.
As discussed before, the Bloch states are fermionic eigenstates of the system.
Thus the input to the procedure is a vector of length $L$ specifying the occupations of each Bloch-state at quasi-momentum $k$.
This determines the electronic part $\ket{\psi_{\mathrm{T}}}_f$ of the trial wavefunction $\ket{\Psi_{\mathrm{T}}} =  \ket{\phi_{\mathrm{T}}}_{b} \otimes \ket{\psi_{\mathrm{T}}}_{f}$, with which we calculate the eigenvalues of the operators $\mathcal{T}$ and $\mathcal{J}$
\begin{equation}
T_{\psi_{\mathrm{T}}} =\, _f\bra{\psi_{\mathrm{T}}} \mathcal{T} \ket{\psi_{\mathrm{T}}}_f \hspace{2mm}; \hspace{2mm} J_{\psi_{\mathrm{T}}} =\, _f\bra{\psi_{\mathrm{T}}}\mathcal{J}\ket{\psi_{\mathrm{T}}}_f.
\end{equation} 
Evaluating the electronic part of the expectation value for the GS energy one is left with the purely photonic Hamiltonian
\begin{equation}
H_{\psi_{\mathrm{T}}} = \omega_0 \left(\ada + \frac{1}{2}\right) + \cos \left(\frac{g}{\sqrt{L}}\left(\aplusa\right)\right) T_{\psi_{\mathrm{T}}}%
+ \sin \left(\frac{g}{\sqrt{L}}\left(\aplusa\right)\right) J_{\psi_{\mathrm{T}}}.
\label{eq:HTrialBosonic}
\end{equation}
The problem reduces to that of  an anharmonic oscillator, that can be solved by numerical diagonalization introducing a cutoff $N_{\rm max}^{\rm boson}$ in the Fock space.
All results are converged with respect to this cutoff.
The scheme then varies over trial wave-functions optimizing for the smallest GS energy of the remaining bosonic problem Eq.~(\ref{eq:HTrialBosonic}).
It thus only compares eigenenergies of exact eigenstates making it possible to find the true GS.
We have chosen different starting wave-functions for the optimization procedure including the state where $\langle \rho_k \rangle = 0.5$ for all $k$ in the BZ and randomly generated states.
Due to somewhat better convergence properties the former have been used to obtain the shown plots.

We verified our results against an exact diagonalization of the full Hamiltonian for small system sizes obtaining identical results within machine precision.

\section*{Data availability}
Data included in the paper can be reproduced using the Python code available at \url{https://github.com/ce335805/comeChainComeShine.git}.

\section*{Code availability}
The code used within this work is openly available at \url{https://github.com/ce335805/comeChainComeShine.git}.


\bibliography{mergedBib.bib}

\begin{thebibliography}{100}
\expandafter\ifx\csname url\endcsname\relax
  \def\url#1{\texttt{#1}}\fi
\expandafter\ifx\csname urlprefix\endcsname\relax\def\urlprefix{URL }\fi
\providecommand{\bibinfo}[2]{#2}
\providecommand{\eprint}[2][]{\url{#2}}

\bibitem{colloquium}
\bibinfo{author}{de~la Torre, A.} \emph{et~al.}
\newblock \bibinfo{title}{Colloquium: Nonthermal pathways to ultrafast control
  in quantum materials}.
\newblock \emph{\bibinfo{journal}{Rev. Mod. Phys.}}
  \textbf{\bibinfo{volume}{93}}, \bibinfo{pages}{041002}
  (\bibinfo{year}{2021}).
\newblock
  \urlprefix\url{https://link.aps.org/doi/10.1103/RevModPhys.93.041002}.

\bibitem{Ac_n_2018}
\bibinfo{author}{Ac{\'{\i}}n, A.} \emph{et~al.}
\newblock \bibinfo{title}{The quantum technologies roadmap: a european
  community view}.
\newblock \emph{\bibinfo{journal}{New Journal of Physics}}
  \textbf{\bibinfo{volume}{20}}, \bibinfo{pages}{080201}
  (\bibinfo{year}{2018}).
\newblock \urlprefix\url{https://doi.org/10.1088/1367-2630/aad1ea}.

\bibitem{Moody2021}
\bibinfo{author}{Moody, G.} \emph{et~al.}
\newblock \bibinfo{title}{2022 roadmap on integrated quantum photonics}.
\newblock \emph{\bibinfo{journal}{Journal of Physics: Photonics}}
  \textbf{\bibinfo{volume}{4}}, \bibinfo{pages}{012501} (\bibinfo{year}{2022}).
\newblock \urlprefix\url{https://doi.org/10.1088/2515-7647/ac1ef4}.

\bibitem{Ebbesen2016}
\bibinfo{author}{Ebbesen, T.~W.}
\newblock \bibinfo{title}{Hybrid light--matter states in a molecular and
  material science perspective}.
\newblock \emph{\bibinfo{journal}{Accounts of Chemical Research}}
  \textbf{\bibinfo{volume}{49}}, \bibinfo{pages}{2403--2412}
  (\bibinfo{year}{2016}).
\newblock \urlprefix\url{https://doi.org/10.1021/acs.accounts.6b00295}.

\bibitem{feist_polaritonic_2018}
\bibinfo{author}{Feist, J.}, \bibinfo{author}{Galego, J.} \&
  \bibinfo{author}{{Garcia-Vidal}, F.~J.}
\newblock \bibinfo{title}{Polaritonic {{Chemistry}} with {{Organic
  Molecules}}}.
\newblock \emph{\bibinfo{journal}{ACS Photonics}} \textbf{\bibinfo{volume}{5}},
  \bibinfo{pages}{205--216} (\bibinfo{year}{2018}).

\bibitem{Ruggenthaler2018}
\bibinfo{author}{Ruggenthaler, M.}, \bibinfo{author}{Tancogne-Dejean, N.},
  \bibinfo{author}{Flick, J.}, \bibinfo{author}{Appel, H.} \&
  \bibinfo{author}{Rubio, A.}
\newblock \bibinfo{title}{From a quantum-electrodynamical light--matter
  description to novel spectroscopies}.
\newblock \emph{\bibinfo{journal}{Nature Reviews Chemistry}}
  \textbf{\bibinfo{volume}{2}}, \bibinfo{pages}{0118} (\bibinfo{year}{2018}).
\newblock \urlprefix\url{https://doi.org/10.1038/s41570-018-0118}.

\bibitem{ribeiro_polariton_2018}
\bibinfo{author}{Ribeiro, R.~F.}, \bibinfo{author}{{Mart{\'i}nez-Mart{\'i}nez},
  L.~A.}, \bibinfo{author}{Du, M.}, \bibinfo{author}{{Campos-Gonzalez-Angulo},
  J.} \& \bibinfo{author}{{Yuen-Zhou}, J.}
\newblock \bibinfo{title}{Polariton chemistry: Controlling molecular dynamics
  with optical cavities}.
\newblock \emph{\bibinfo{journal}{Chemical Science}}
  \textbf{\bibinfo{volume}{9}}, \bibinfo{pages}{6325--6339}
  (\bibinfo{year}{2018}).

\bibitem{flick_strong_2018}
\bibinfo{author}{Flick, J.}, \bibinfo{author}{Rivera, N.} \&
  \bibinfo{author}{Narang, P.}
\newblock \bibinfo{title}{Strong light-matter coupling in quantum chemistry and
  quantum photonics}.
\newblock \emph{\bibinfo{journal}{Nanophotonics}} \textbf{\bibinfo{volume}{7}},
  \bibinfo{pages}{1479--1501} (\bibinfo{year}{2018}).
\newblock
  \urlprefix\url{https://www.degruyter.com/view/j/nanoph.2018.7.issue-9/nanoph-2018-0067/nanoph-2018-0067.xml}.

\bibitem{FriskKockum2019}
\bibinfo{author}{Frisk~Kockum, A.}, \bibinfo{author}{Miranowicz, A.},
  \bibinfo{author}{De~Liberato, S.}, \bibinfo{author}{Savasta, S.} \&
  \bibinfo{author}{Nori, F.}
\newblock \bibinfo{title}{Ultrastrong coupling between light and matter}.
\newblock \emph{\bibinfo{journal}{Nature Reviews Physics}}
  \textbf{\bibinfo{volume}{1}}, \bibinfo{pages}{19--40} (\bibinfo{year}{2019}).
\newblock \urlprefix\url{https://doi.org/10.1038/s42254-018-0006-2}.

\bibitem{wang_observation_2013}
\bibinfo{author}{Wang, Y.~H.}, \bibinfo{author}{Steinberg, H.},
  \bibinfo{author}{Jarillo-Herrero, P.} \& \bibinfo{author}{Gedik, N.}
\newblock \bibinfo{title}{Observation of {F}loquet-{B}loch states on the
  surface of a {T}opological {I}nsulator}.
\newblock \emph{\bibinfo{journal}{Science}} \textbf{\bibinfo{volume}{342}},
  \bibinfo{pages}{453--457} (\bibinfo{year}{2013}).
\newblock \urlprefix\url{https://science.sciencemag.org/content/342/6157/453}.

\bibitem{mciver_light-induced_2020}
\bibinfo{author}{McIver, J.~W.} \emph{et~al.}
\newblock \bibinfo{title}{Light-induced anomalous {Hall} effect in graphene}.
\newblock \emph{\bibinfo{journal}{Nature Physics}}
  \textbf{\bibinfo{volume}{16}}, \bibinfo{pages}{38--41}
  (\bibinfo{year}{2020}).
\newblock \urlprefix\url{https://www.nature.com/articles/s41567-019-0698-y}.

\bibitem{bukov_universal_2015}
\bibinfo{author}{Bukov, M.}, \bibinfo{author}{D'Alessio, L.} \&
  \bibinfo{author}{Polkovnikov, A.}
\newblock \bibinfo{title}{Universal high-frequency behavior of periodically
  driven systems: from dynamical stabilization to {Floquet} engineering}.
\newblock \emph{\bibinfo{journal}{Advances in Physics}}
  \textbf{\bibinfo{volume}{64}}, \bibinfo{pages}{139--226}
  (\bibinfo{year}{2015}).
\newblock \urlprefix\url{http://dx.doi.org/10.1080/00018732.2015.1055918}.

\bibitem{Eckardt2017}
\bibinfo{author}{Eckardt, A.}
\newblock \bibinfo{title}{{Colloquium: Atomic quantum gases in periodically
  driven optical lattices}}.
\newblock \emph{\bibinfo{journal}{Reviews of Modern Physics}}
  \textbf{\bibinfo{volume}{89}}, \bibinfo{pages}{011004}
  (\bibinfo{year}{2017}).
\newblock \urlprefix\url{http://link.aps.org/doi/10.1103/RevModPhys.89.011004}.
\newblock \eprint{1606.08041}.

\bibitem{oka_floquet_2019}
\bibinfo{author}{Oka, T.} \& \bibinfo{author}{Kitamura, S.}
\newblock \bibinfo{title}{Floquet {Engineering} of {Quantum} {Materials}}.
\newblock \emph{\bibinfo{journal}{Annual Review of Condensed Matter Physics}}
  \textbf{\bibinfo{volume}{10}}, \bibinfo{pages}{387--408}
  (\bibinfo{year}{2019}).
\newblock
  \urlprefix\url{https://doi.org/10.1146/annurev-conmatphys-031218-013423}.
\newblock \bibinfo{note}{\_eprint:
  https://doi.org/10.1146/annurev-conmatphys-031218-013423}.

\bibitem{Rudner_2020}
\bibinfo{author}{Rudner, M.~S.} \& \bibinfo{author}{Lindner, N.~H.}
\newblock \bibinfo{title}{The floquet engineer's handbook}
  (\bibinfo{year}{2020}).
\newblock \eprint{arXiv:2003.08252}.

\bibitem{OkaAoki}
\bibinfo{author}{{Oka}, T.} \& \bibinfo{author}{{Aoki}, H.}
\newblock \bibinfo{title}{{Photovoltaic Hall effect in graphene}}.
\newblock \emph{\bibinfo{journal}{Physical Review B}}
  \textbf{\bibinfo{volume}{79}}, \bibinfo{pages}{081406}
  (\bibinfo{year}{2009}).

\bibitem{Lindner2011}
\bibinfo{author}{Lindner, N.~H.}, \bibinfo{author}{Refael, G.} \&
  \bibinfo{author}{Galitski, V.}
\newblock \bibinfo{title}{Floquet topological insulator in semiconductor
  quantum wells}.
\newblock \emph{\bibinfo{journal}{Nature Physics}}
  \textbf{\bibinfo{volume}{7}}, \bibinfo{pages}{490--495}
  (\bibinfo{year}{2011}).
\newblock \urlprefix\url{https://doi.org/10.1038/nphys1926}.

\bibitem{Kitagawa2011}
\bibinfo{author}{Kitagawa, T.}, \bibinfo{author}{Oka, T.},
  \bibinfo{author}{Brataas, A.}, \bibinfo{author}{Fu, L.} \&
  \bibinfo{author}{Demler, E.}
\newblock \bibinfo{title}{Transport properties of nonequilibrium systems under
  the application of light: Photoinduced quantum hall insulators without landau
  levels}.
\newblock \emph{\bibinfo{journal}{Phys. Rev. B}} \textbf{\bibinfo{volume}{84}},
  \bibinfo{pages}{235108} (\bibinfo{year}{2011}).
\newblock \urlprefix\url{https://link.aps.org/doi/10.1103/PhysRevB.84.235108}.

\bibitem{Decker_2019}
\bibinfo{author}{Decker, K. S.~C.}, \bibinfo{author}{Karrasch, C.},
  \bibinfo{author}{Eisert, J.} \& \bibinfo{author}{Kennes, D.~M.}
\newblock \bibinfo{title}{Floquet engineering topological many-body localized
  systems}.
\newblock \emph{\bibinfo{journal}{Phys. Rev. Lett.}}
  \textbf{\bibinfo{volume}{124}}, \bibinfo{pages}{190601}
  (\bibinfo{year}{2020}).
\newblock
  \urlprefix\url{https://link.aps.org/doi/10.1103/PhysRevLett.124.190601}.

\bibitem{sentef_theory_2015}
\bibinfo{author}{Sentef, M.~A.} \emph{et~al.}
\newblock \bibinfo{title}{Theory of {Floquet} band formation and local
  pseudospin textures in pump-probe photoemission of graphene}.
\newblock \emph{\bibinfo{journal}{Nature Communications}}
  \textbf{\bibinfo{volume}{6}}, \bibinfo{pages}{7047} (\bibinfo{year}{2015}).
\newblock \urlprefix\url{https://www.nature.com/articles/ncomms8047}.

\bibitem{Hubener2017}
\bibinfo{author}{H{\"{u}}bener, H.}, \bibinfo{author}{Sentef, M.~A.},
  \bibinfo{author}{{De Giovannini}, U.}, \bibinfo{author}{Kemper, A.~F.} \&
  \bibinfo{author}{Rubio, A.}
\newblock \bibinfo{title}{{Creating stable Floquet-Weyl semimetals by
  laser-driving of 3D Dirac materials}}.
\newblock \emph{\bibinfo{journal}{Nature Communications}}
  \textbf{\bibinfo{volume}{8}}, \bibinfo{pages}{13940} (\bibinfo{year}{2017}).
\newblock \eprint{1604.03399}.

\bibitem{Fleckenstein2020}
\bibinfo{author}{Fleckenstein, C.}, \bibinfo{author}{Ziani, N.~T.},
  \bibinfo{author}{Privitera, L.}, \bibinfo{author}{Sassetti, M.} \&
  \bibinfo{author}{Trauzettel, B.}
\newblock \bibinfo{title}{Transport signatures of a floquet topological
  transition at the helical edge}.
\newblock \emph{\bibinfo{journal}{Phys. Rev. B}}
  \textbf{\bibinfo{volume}{101}}, \bibinfo{pages}{201401}
  (\bibinfo{year}{2020}).
\newblock \urlprefix\url{https://link.aps.org/doi/10.1103/PhysRevB.101.201401}.

\bibitem{Bukov2016}
\bibinfo{author}{Bukov, M.}, \bibinfo{author}{Kolodrubetz, M.} \&
  \bibinfo{author}{Polkovnikov, A.}
\newblock \bibinfo{title}{Schrieffer-wolff transformation for periodically
  driven systems: Strongly correlated systems with artificial gauge fields}.
\newblock \emph{\bibinfo{journal}{Phys. Rev. Lett.}}
  \textbf{\bibinfo{volume}{116}}, \bibinfo{pages}{125301}
  (\bibinfo{year}{2016}).
\newblock
  \urlprefix\url{https://link.aps.org/doi/10.1103/PhysRevLett.116.125301}.

\bibitem{Claassen2017}
\bibinfo{author}{Claassen, M.}, \bibinfo{author}{Jiang, H.~C.},
  \bibinfo{author}{Moritz, B.} \& \bibinfo{author}{Devereaux, T.~P.}
\newblock \bibinfo{title}{{Dynamical time-reversal symmetry breaking and
  photo-induced chiral spin liquids in frustrated Mott insulators}}.
\newblock \emph{\bibinfo{journal}{Nature Communications}}
  \textbf{\bibinfo{volume}{8}}, \bibinfo{pages}{1192} (\bibinfo{year}{2017}).
\newblock
  \urlprefix\url{https://www.nature.com/articles/s41467-017-00876-y.pdf}.
\newblock \eprint{1611.07964}.

\bibitem{Kennes2018-FloquetChains}
\bibinfo{author}{Kennes, D.~M.}, \bibinfo{author}{de~la Torre, A.},
  \bibinfo{author}{Ron, A.}, \bibinfo{author}{Hsieh, D.} \&
  \bibinfo{author}{Millis, A.~J.}
\newblock \bibinfo{title}{Floquet engineering in quantum chains}.
\newblock \emph{\bibinfo{journal}{Phys. Rev. Lett.}}
  \textbf{\bibinfo{volume}{120}}, \bibinfo{pages}{127601}
  (\bibinfo{year}{2018}).
\newblock
  \urlprefix\url{https://link.aps.org/doi/10.1103/PhysRevLett.120.127601}.

\bibitem{Mentink2015}
\bibinfo{author}{Mentink, J.~H.}, \bibinfo{author}{Balzer, K.} \&
  \bibinfo{author}{Eckstein, M.}
\newblock \bibinfo{title}{{Ultrafast and reversible control of the exchange
  interaction in {M}ott insulators}}.
\newblock \emph{\bibinfo{journal}{Nature Communications}}
  \textbf{\bibinfo{volume}{6}}, \bibinfo{pages}{6708} (\bibinfo{year}{2015}).
\newblock \urlprefix\url{https://www.nature.com/articles/ncomms7708.pdf}.
\newblock \eprint{arXiv:1407.4761v1}.

\bibitem{Walldorf2019}
\bibinfo{author}{Walldorf, N.}, \bibinfo{author}{Kennes, D.~M.},
  \bibinfo{author}{Paaske, J.} \& \bibinfo{author}{Millis, A.~J.}
\newblock \bibinfo{title}{The antiferromagnetic phase of the floquet-driven
  hubbard model}.
\newblock \emph{\bibinfo{journal}{Phys. Rev. B}}
  \textbf{\bibinfo{volume}{100}}, \bibinfo{pages}{121110}
  (\bibinfo{year}{2019}).
\newblock \urlprefix\url{https://link.aps.org/doi/10.1103/PhysRevB.100.121110}.

\bibitem{Sentef2016}
\bibinfo{author}{Sentef, M.~A.}, \bibinfo{author}{Kemper, A.~F.},
  \bibinfo{author}{Georges, A.} \& \bibinfo{author}{Kollath, C.}
\newblock \bibinfo{title}{{Theory of light-enhanced phonon-mediated
  superconductivity}}.
\newblock \emph{\bibinfo{journal}{Physical Review B}}
  \textbf{\bibinfo{volume}{93}}, \bibinfo{pages}{1--10} (\bibinfo{year}{2016}).
\newblock \eprint{1505.07575}.

\bibitem{Knap2016}
\bibinfo{author}{Knap, M.}, \bibinfo{author}{Babadi, M.},
  \bibinfo{author}{Refael, G.}, \bibinfo{author}{Martin, I.} \&
  \bibinfo{author}{Demler, E.}
\newblock \bibinfo{title}{{Dynamical Cooper pairing in nonequilibrium
  electron-phonon systems}}.
\newblock \emph{\bibinfo{journal}{Physical Review B}}
  \textbf{\bibinfo{volume}{94}}, \bibinfo{pages}{214504}
  (\bibinfo{year}{2016}).
\newblock \eprint{1511.07874}.

\bibitem{Kennes2017}
\bibinfo{author}{Kennes, D.~M.}, \bibinfo{author}{Wilner, E.~Y.},
  \bibinfo{author}{Reichman, D.~R.} \& \bibinfo{author}{Millis, A.~J.}
\newblock \bibinfo{title}{Transient superconductivity from electronic squeezing
  of optically pumped phonons}.
\newblock \emph{\bibinfo{journal}{Nature Physics}}
  \textbf{\bibinfo{volume}{13}}, \bibinfo{pages}{479--483}
  (\bibinfo{year}{2017}).
\newblock \eprint{1609.03802v1}.

\bibitem{Murakami2017}
\bibinfo{author}{Murakami, Y.}, \bibinfo{author}{Tsuji, N.},
  \bibinfo{author}{Eckstein, M.} \& \bibinfo{author}{Werner, P.}
\newblock \bibinfo{title}{{Nonequilibrium steady states and transient dynamics
  of conventional superconductors under phonon driving}}.
\newblock \emph{\bibinfo{journal}{Physical Review B}}
  \textbf{\bibinfo{volume}{96}}, \bibinfo{pages}{045125}
  (\bibinfo{year}{2017}).
\newblock \eprint{1702.02942}.

\bibitem{Porta2019}
\bibinfo{author}{Porta, S.} \emph{et~al.}
\newblock \bibinfo{title}{Feasible model for photoinduced interband pairing}.
\newblock \emph{\bibinfo{journal}{Phys. Rev. B}}
  \textbf{\bibinfo{volume}{100}}, \bibinfo{pages}{024513}
  (\bibinfo{year}{2019}).
\newblock \urlprefix\url{https://link.aps.org/doi/10.1103/PhysRevB.100.024513}.

\bibitem{Kennes_2018}
\bibinfo{author}{Kennes, D.~M.}, \bibinfo{author}{Claassen, M.},
  \bibinfo{author}{Sentef, M.~A.} \& \bibinfo{author}{Karrasch, C.}
\newblock \bibinfo{title}{Light-induced $d$-wave superconductivity through
  floquet-engineered fermi surfaces in cuprates}.
\newblock \emph{\bibinfo{journal}{Phys. Rev. B}}
  \textbf{\bibinfo{volume}{100}}, \bibinfo{pages}{075115}
  (\bibinfo{year}{2019}).
\newblock \urlprefix\url{https://link.aps.org/doi/10.1103/PhysRevB.100.075115}.

\bibitem{DAlessio2014}
\bibinfo{author}{D'Alessio, L.} \& \bibinfo{author}{Rigol, M.}
\newblock \bibinfo{title}{Long-time behavior of isolated periodically driven
  interacting lattice systems}.
\newblock \emph{\bibinfo{journal}{Phys. Rev. X}} \textbf{\bibinfo{volume}{4}},
  \bibinfo{pages}{041048} (\bibinfo{year}{2014}).
\newblock \urlprefix\url{https://link.aps.org/doi/10.1103/PhysRevX.4.041048}.

\bibitem{Lazarides2014}
\bibinfo{author}{Lazarides, A.}, \bibinfo{author}{Das, A.} \&
  \bibinfo{author}{Moessner, R.}
\newblock \bibinfo{title}{Equilibrium states of generic quantum systems subject
  to periodic driving}.
\newblock \emph{\bibinfo{journal}{Phys. Rev. E}} \textbf{\bibinfo{volume}{90}},
  \bibinfo{pages}{012110} (\bibinfo{year}{2014}).
\newblock \urlprefix\url{https://link.aps.org/doi/10.1103/PhysRevE.90.012110}.

\bibitem{Kibis2011}
\bibinfo{author}{Kibis, O.~V.}, \bibinfo{author}{Kyriienko, O.} \&
  \bibinfo{author}{Shelykh, I.~A.}
\newblock \bibinfo{title}{Band gap in graphene induced by vacuum fluctuations}.
\newblock \emph{\bibinfo{journal}{Phys. Rev. B}} \textbf{\bibinfo{volume}{84}},
  \bibinfo{pages}{195413} (\bibinfo{year}{2011}).
\newblock \urlprefix\url{https://link.aps.org/doi/10.1103/PhysRevB.84.195413}.

\bibitem{wang_cavity_2019}
\bibinfo{author}{Wang, X.}, \bibinfo{author}{Ronca, E.} \&
  \bibinfo{author}{Sentef, M.~A.}
\newblock \bibinfo{title}{Cavity quantum electrodynamical {Chern} insulator:
  {Towards} light-induced quantized anomalous {Hall} effect in graphene}.
\newblock \emph{\bibinfo{journal}{Physical Review B}}
  \textbf{\bibinfo{volume}{99}}, \bibinfo{pages}{235156}
  (\bibinfo{year}{2019}).
\newblock \urlprefix\url{https://link.aps.org/doi/10.1103/PhysRevB.99.235156}.

\bibitem{Huebener2021}
\bibinfo{author}{H{\"u}bener, H.} \emph{et~al.}
\newblock \bibinfo{title}{Engineering quantum materials with chiral optical
  cavities}.
\newblock \emph{\bibinfo{journal}{Nature Materials}}
  \textbf{\bibinfo{volume}{20}}, \bibinfo{pages}{438--442}
  (\bibinfo{year}{2021}).
\newblock \urlprefix\url{https://doi.org/10.1038/s41563-020-00801-7}.

\bibitem{Dutra2004}
\bibinfo{author}{Dutra, S.~M.}
\newblock \emph{\bibinfo{title}{Cavity Quantum Electrodynamics}}
  (\bibinfo{publisher}{John Wiley {\&} Sons, Inc.}, \bibinfo{year}{2004}).
\newblock \urlprefix\url{https://doi.org/10.1002/0471713465}.

\bibitem{Li2020-Quantization}
\bibinfo{author}{Li, J.} \emph{et~al.}
\newblock \bibinfo{title}{Electromagnetic coupling in tight-binding models for
  strongly correlated light and matter}.
\newblock \emph{\bibinfo{journal}{Phys. Rev. B}}
  \textbf{\bibinfo{volume}{101}}, \bibinfo{pages}{205140}
  (\bibinfo{year}{2020}).
\newblock \urlprefix\url{https://link.aps.org/doi/10.1103/PhysRevB.101.205140}.

\bibitem{Maissen2014}
\bibinfo{author}{Maissen, C.} \emph{et~al.}
\newblock \bibinfo{title}{Ultrastrong coupling in the near field of
  complementary split-ring resonators}.
\newblock \emph{\bibinfo{journal}{Phys. Rev. B}} \textbf{\bibinfo{volume}{90}},
  \bibinfo{pages}{205309} (\bibinfo{year}{2014}).
\newblock \urlprefix\url{https://link.aps.org/doi/10.1103/PhysRevB.90.205309}.

\bibitem{Meschede1985}
\bibinfo{author}{Meschede, D.}, \bibinfo{author}{Walther, H.} \&
  \bibinfo{author}{M\"uller, G.}
\newblock \bibinfo{title}{One-atom maser}.
\newblock \emph{\bibinfo{journal}{Phys. Rev. Lett.}}
  \textbf{\bibinfo{volume}{54}}, \bibinfo{pages}{551--554}
  (\bibinfo{year}{1985}).
\newblock \urlprefix\url{https://link.aps.org/doi/10.1103/PhysRevLett.54.551}.

\bibitem{Thompson1992}
\bibinfo{author}{Thompson, R.~J.}, \bibinfo{author}{Rempe, G.} \&
  \bibinfo{author}{Kimble, H.~J.}
\newblock \bibinfo{title}{Observation of normal-mode splitting for an atom in
  an optical cavity}.
\newblock \emph{\bibinfo{journal}{Phys. Rev. Lett.}}
  \textbf{\bibinfo{volume}{68}}, \bibinfo{pages}{1132--1135}
  (\bibinfo{year}{1992}).
\newblock \urlprefix\url{https://link.aps.org/doi/10.1103/PhysRevLett.68.1132}.

\bibitem{GU20171}
\bibinfo{author}{Gu, X.}, \bibinfo{author}{Kockum, A.~F.},
  \bibinfo{author}{Miranowicz, A.}, \bibinfo{author}{xi~Liu, Y.} \&
  \bibinfo{author}{Nori, F.}
\newblock \bibinfo{title}{Microwave photonics with superconducting quantum
  circuits}.
\newblock \emph{\bibinfo{journal}{Physics Reports}}
  \textbf{\bibinfo{volume}{718-719}}, \bibinfo{pages}{1--102}
  (\bibinfo{year}{2017}).
\newblock
  \urlprefix\url{https://www.sciencedirect.com/science/article/pii/S0370157317303290}.
\newblock \bibinfo{note}{Microwave photonics with superconducting quantum
  circuits}.

\bibitem{Scalari2012}
\bibinfo{author}{Scalari, G.} \emph{et~al.}
\newblock \bibinfo{title}{Ultrastrong coupling of the cyclotron transition of a
  2d electron gas to a {THz} metamaterial}.
\newblock \emph{\bibinfo{journal}{Science}} \textbf{\bibinfo{volume}{335}},
  \bibinfo{pages}{1323--1326} (\bibinfo{year}{2012}).
\newblock \urlprefix\url{https://doi.org/10.1126/science.1216022}.

\bibitem{Keller2017}
\bibinfo{author}{Keller, J.} \emph{et~al.}
\newblock \bibinfo{title}{Few-electron ultrastrong light-matter coupling at 300
  ghz with nanogap hybrid lc microcavities}.
\newblock \emph{\bibinfo{journal}{Nano Letters}} \textbf{\bibinfo{volume}{17}},
  \bibinfo{pages}{7410--7415} (\bibinfo{year}{2017}).
\newblock \urlprefix\url{https://doi.org/10.1021/acs.nanolett.7b03228}.

\bibitem{BallariniDeLiberato_2019}
\bibinfo{author}{Ballarini, D.} \& \bibinfo{author}{Liberato, S.~D.}
\newblock \bibinfo{title}{Polaritonics: from microcavities to sub-wavelength
  confinement}.
\newblock \emph{\bibinfo{journal}{Nanophotonics}} \textbf{\bibinfo{volume}{8}},
  \bibinfo{pages}{641--654} (\bibinfo{year}{2019}).
\newblock \urlprefix\url{https://doi.org/10.1515/nanoph-2018-0188}.

\bibitem{ParaviciniBagliani2018}
\bibinfo{author}{Paravicini-Bagliani, G.~L.} \emph{et~al.}
\newblock \bibinfo{title}{Magneto-transport controlled by landau polariton
  states}.
\newblock \emph{\bibinfo{journal}{Nature Physics}}
  \textbf{\bibinfo{volume}{15}}, \bibinfo{pages}{186--190}
  (\bibinfo{year}{2018}).
\newblock \urlprefix\url{https://doi.org/10.1038/s41567-018-0346-y}.

\bibitem{kasprzak_bose-einstein_2006}
\bibinfo{author}{Kasprzak, J.} \emph{et~al.}
\newblock \bibinfo{title}{Bose\textendash{{Einstein}} condensation of exciton
  polaritons}.
\newblock \emph{\bibinfo{journal}{Nature}} \textbf{\bibinfo{volume}{443}},
  \bibinfo{pages}{409--414} (\bibinfo{year}{2006}).

\bibitem{Keeling_2020}
\bibinfo{author}{Keeling, J.} \& \bibinfo{author}{Kéna-Cohen, S.}
\newblock \bibinfo{title}{Bose–einstein condensation of exciton-polaritons in
  organic microcavities}.
\newblock \emph{\bibinfo{journal}{Annual Review of Physical Chemistry}}
  \textbf{\bibinfo{volume}{71}}, \bibinfo{pages}{435--459}
  (\bibinfo{year}{2020}).
\newblock
  \urlprefix\url{https://doi.org/10.1146/annurev-physchem-010920-102509}.
\newblock \bibinfo{note}{PMID: 32126177},
  \eprint{https://doi.org/10.1146/annurev-physchem-010920-102509}.

\bibitem{byrnes_exciton-polariton_2014}
\bibinfo{author}{Byrnes, T.}, \bibinfo{author}{Kim, N.~Y.} \&
  \bibinfo{author}{Yamamoto, Y.}
\newblock \bibinfo{title}{Exciton-polariton condensates}.
\newblock \emph{\bibinfo{journal}{Nature Physics}}
  \textbf{\bibinfo{volume}{10}}, \bibinfo{pages}{803--813}
  (\bibinfo{year}{2014}).

\bibitem{Anoop2016}
\bibinfo{author}{Thomas, A.} \emph{et~al.}
\newblock \bibinfo{title}{Ground-state chemical reactivity under vibrational
  coupling to the vacuum electromagnetic field}.
\newblock \emph{\bibinfo{journal}{Angewandte Chemie International Edition}}
  \textbf{\bibinfo{volume}{55}}, \bibinfo{pages}{11462--11466}
  (\bibinfo{year}{2016}).
\newblock
  \urlprefix\url{https://onlinelibrary.wiley.com/doi/abs/10.1002/anie.201605504}.
\newblock
  \eprint{https://onlinelibrary.wiley.com/doi/pdf/10.1002/anie.201605504}.

\bibitem{Schaefer2021}
\bibinfo{author}{Schäfer, C.}, \bibinfo{author}{Flick, J.},
  \bibinfo{author}{Ronca, E.}, \bibinfo{author}{Narang, P.} \&
  \bibinfo{author}{Rubio, A.}
\newblock \bibinfo{title}{Shining light on the microscopic resonant mechanism
  responsible for cavity-mediated chemical reactivity} (\bibinfo{year}{2021}).
\newblock \eprint{arXiv:2104.12429}.

\bibitem{sentef_cavity_2018}
\bibinfo{author}{Sentef, M.~A.}, \bibinfo{author}{Ruggenthaler, M.} \&
  \bibinfo{author}{Rubio, A.}
\newblock \bibinfo{title}{Cavity quantum-electrodynamical polaritonically
  enhanced electron-phonon coupling and its influence on superconductivity}.
\newblock \emph{\bibinfo{journal}{Science Advances}}
  \textbf{\bibinfo{volume}{4}}, \bibinfo{pages}{eaau6969}
  (\bibinfo{year}{2018}).
\newblock \urlprefix\url{http://advances.sciencemag.org/content/4/11/eaau6969}.

\bibitem{curtis2021}
\bibinfo{author}{Curtis, J.~B.} \emph{et~al.}
\newblock \bibinfo{title}{Cavity magnon-polaritons in cuprate parent
  compounds}.
\newblock \emph{\bibinfo{journal}{Phys. Rev. Research}}
  \textbf{\bibinfo{volume}{4}}, \bibinfo{pages}{013101} (\bibinfo{year}{2022}).
\newblock
  \urlprefix\url{https://link.aps.org/doi/10.1103/PhysRevResearch.4.013101}.

\bibitem{schlawin_cavity-mediated_2019}
\bibinfo{author}{Schlawin, F.}, \bibinfo{author}{Cavalleri, A.} \&
  \bibinfo{author}{Jaksch, D.}
\newblock \bibinfo{title}{Cavity-{Mediated} {Electron}-{Photon}
  {Superconductivity}}.
\newblock \emph{\bibinfo{journal}{Physical Review Letters}}
  \textbf{\bibinfo{volume}{122}}, \bibinfo{pages}{133602}
  (\bibinfo{year}{2019}).
\newblock
  \urlprefix\url{https://link.aps.org/doi/10.1103/PhysRevLett.122.133602}.

\bibitem{chakraborty_non-bcs-type_2020}
\bibinfo{author}{Chakraborty, A.} \& \bibinfo{author}{Piazza, F.}
\newblock \bibinfo{title}{Long-range photon fluctuations enhance
  photon-mediated electron pairing and superconductivity}.
\newblock \emph{\bibinfo{journal}{Phys. Rev. Lett.}}
  \textbf{\bibinfo{volume}{127}}, \bibinfo{pages}{177002}
  (\bibinfo{year}{2021}).
\newblock
  \urlprefix\url{https://link.aps.org/doi/10.1103/PhysRevLett.127.177002}.

\bibitem{Gao_Schlawin_2020}
\bibinfo{author}{Gao, H.}, \bibinfo{author}{Schlawin, F.},
  \bibinfo{author}{Buzzi, M.}, \bibinfo{author}{Cavalleri, A.} \&
  \bibinfo{author}{Jaksch, D.}
\newblock \bibinfo{title}{Photoinduced electron pairing in a driven cavity}.
\newblock \emph{\bibinfo{journal}{Phys. Rev. Lett.}}
  \textbf{\bibinfo{volume}{125}}, \bibinfo{pages}{053602}
  (\bibinfo{year}{2020}).
\newblock
  \urlprefix\url{https://link.aps.org/doi/10.1103/PhysRevLett.125.053602}.

\bibitem{curtis_cavity_2019}
\bibinfo{author}{Curtis, J.~B.}, \bibinfo{author}{Raines, Z.~M.},
  \bibinfo{author}{Allocca, A.~A.}, \bibinfo{author}{Hafezi, M.} \&
  \bibinfo{author}{Galitski, V.~M.}
\newblock \bibinfo{title}{Cavity {Quantum} {Eliashberg} {Enhancement} of
  {Superconductivity}}.
\newblock \emph{\bibinfo{journal}{Physical Review Letters}}
  \textbf{\bibinfo{volume}{122}}, \bibinfo{pages}{167002}
  (\bibinfo{year}{2019}).
\newblock
  \urlprefix\url{https://link.aps.org/doi/10.1103/PhysRevLett.122.167002}.

\bibitem{Allocca2019}
\bibinfo{author}{Allocca, A.~A.}, \bibinfo{author}{Raines, Z.~M.},
  \bibinfo{author}{Curtis, J.~B.} \& \bibinfo{author}{Galitski, V.~M.}
\newblock \bibinfo{title}{Cavity superconductor-polaritons}.
\newblock \emph{\bibinfo{journal}{Phys. Rev. B}} \textbf{\bibinfo{volume}{99}},
  \bibinfo{pages}{020504} (\bibinfo{year}{2019}).
\newblock \urlprefix\url{https://link.aps.org/doi/10.1103/PhysRevB.99.020504}.

\bibitem{thomas_exploring_2019}
\bibinfo{author}{Thomas, A.} \emph{et~al.}
\newblock \bibinfo{title}{Exploring {Superconductivity} under {Strong}
  {Coupling} with the {Vacuum} {Electromagnetic} {Field}}.
\newblock \emph{\bibinfo{journal}{arXiv:1911.01459 [cond-mat,
  physics:quant-ph]}}  (\bibinfo{year}{2019}).
\newblock \urlprefix\url{http://arxiv.org/abs/1911.01459}.
\newblock \bibinfo{note}{ArXiv: 1911.01459}.

\bibitem{Nataf2010}
\bibinfo{author}{Nataf, P.} \& \bibinfo{author}{Ciuti, C.}
\newblock \bibinfo{title}{No-go theorem for superradiant quantum phase
  transitions in cavity qed and counter-example in circuit qed}.
\newblock \emph{\bibinfo{journal}{Nature Communications}}
  \textbf{\bibinfo{volume}{1}}, \bibinfo{pages}{72} (\bibinfo{year}{2010}).
\newblock \urlprefix\url{https://doi.org/10.1038/ncomms1069}.

\bibitem{mazza_superradiant_2019}
\bibinfo{author}{Mazza, G.} \& \bibinfo{author}{Georges, A.}
\newblock \bibinfo{title}{Superradiant {Quantum} {Materials}}.
\newblock \emph{\bibinfo{journal}{Physical Review Letters}}
  \textbf{\bibinfo{volume}{122}}, \bibinfo{pages}{017401}
  (\bibinfo{year}{2019}).
\newblock
  \urlprefix\url{https://link.aps.org/doi/10.1103/PhysRevLett.122.017401}.

\bibitem{andolina_cavity_2019}
\bibinfo{author}{Andolina, G.~M.}, \bibinfo{author}{Pellegrino, F. M.~D.},
  \bibinfo{author}{Giovannetti, V.}, \bibinfo{author}{MacDonald, A.~H.} \&
  \bibinfo{author}{Polini, M.}
\newblock \bibinfo{title}{Cavity quantum electrodynamics of strongly correlated
  electron systems: {A} no-go theorem for photon condensation}.
\newblock \emph{\bibinfo{journal}{Physical Review B}}
  \textbf{\bibinfo{volume}{100}}, \bibinfo{pages}{121109}
  (\bibinfo{year}{2019}).
\newblock \urlprefix\url{https://link.aps.org/doi/10.1103/PhysRevB.100.121109}.

\bibitem{ashida_demler}
\bibinfo{author}{Ashida, Y.}, \bibinfo{author}{Imamoglu, A.} \&
  \bibinfo{author}{Demler, E.}
\newblock \bibinfo{title}{Nonperturbative waveguide quantum electrodynamics}
  (\bibinfo{year}{2021}).
\newblock \eprint{arXiv:2105.08833}.

\bibitem{Schuler2020}
\bibinfo{author}{Schuler, M.}, \bibinfo{author}{Bernardis, D.~D.},
  \bibinfo{author}{Läuchli, A.~M.} \& \bibinfo{author}{Rabl, P.}
\newblock \bibinfo{title}{{The Vacua of Dipolar Cavity Quantum
  Electrodynamics}}.
\newblock \emph{\bibinfo{journal}{SciPost Phys.}} \textbf{\bibinfo{volume}{9}},
  \bibinfo{pages}{66} (\bibinfo{year}{2020}).
\newblock \urlprefix\url{https://scipost.org/10.21468/SciPostPhys.9.5.066}.

\bibitem{Bernardis2018}
\bibinfo{author}{De~Bernardis, D.}, \bibinfo{author}{Jaako, T.} \&
  \bibinfo{author}{Rabl, P.}
\newblock \bibinfo{title}{Cavity quantum electrodynamics in the nonperturbative
  regime}.
\newblock \emph{\bibinfo{journal}{Phys. Rev. A}} \textbf{\bibinfo{volume}{97}},
  \bibinfo{pages}{043820} (\bibinfo{year}{2018}).
\newblock \urlprefix\url{https://link.aps.org/doi/10.1103/PhysRevA.97.043820}.

\bibitem{Guerci2020}
\bibinfo{author}{Guerci, D.}, \bibinfo{author}{Simon, P.} \&
  \bibinfo{author}{Mora, C.}
\newblock \bibinfo{title}{Superradiant phase transition in electronic systems
  and emergent topological phases}.
\newblock \emph{\bibinfo{journal}{Phys. Rev. Lett.}}
  \textbf{\bibinfo{volume}{125}}, \bibinfo{pages}{257604}
  (\bibinfo{year}{2020}).
\newblock
  \urlprefix\url{https://link.aps.org/doi/10.1103/PhysRevLett.125.257604}.

\bibitem{Reitz2021}
\bibinfo{author}{Reitz, M.}, \bibinfo{author}{Sommer, C.} \&
  \bibinfo{author}{Genes, C.}
\newblock \bibinfo{title}{Cooperative quantum phenomena in light-matter
  platforms}.
\newblock \emph{\bibinfo{journal}{PRX Quantum}} \textbf{\bibinfo{volume}{3}},
  \bibinfo{pages}{010201} (\bibinfo{year}{2022}).
\newblock \urlprefix\url{https://link.aps.org/doi/10.1103/PRXQuantum.3.010201}.

\bibitem{stokes_uniqueness_2020}
\bibinfo{author}{Stokes, A.} \& \bibinfo{author}{Nazir, A.}
\newblock \bibinfo{title}{Uniqueness of the {Phase} {Transition} in
  {Many}-{Dipole} {Cavity} {Quantum} {Electrodynamical} {Systems}}.
\newblock \emph{\bibinfo{journal}{Phys. Rev. Lett.}}
  \textbf{\bibinfo{volume}{125}}, \bibinfo{pages}{143603}
  (\bibinfo{year}{2020}).
\newblock
  \urlprefix\url{https://link.aps.org/doi/10.1103/PhysRevLett.125.143603}.
\newblock \bibinfo{note}{Publisher: American Physical Society}.

\bibitem{Genet2021}
\bibinfo{author}{Genet, C.}, \bibinfo{author}{Faist, J.} \&
  \bibinfo{author}{Ebbesen, T.~W.}
\newblock \bibinfo{title}{Inducing new material properties with hybrid
  light{\textendash}matter states}.
\newblock \emph{\bibinfo{journal}{Physics Today}}
  \textbf{\bibinfo{volume}{74}}, \bibinfo{pages}{42--48}
  (\bibinfo{year}{2021}).
\newblock \urlprefix\url{https://doi.org/10.1063/pt.3.4749}.

\bibitem{Dicke1954}
\bibinfo{author}{Dicke, R.~H.}
\newblock \bibinfo{title}{Coherence in spontaneous radiation processes}.
\newblock \emph{\bibinfo{journal}{Phys. Rev.}} \textbf{\bibinfo{volume}{93}},
  \bibinfo{pages}{99--110} (\bibinfo{year}{1954}).
\newblock \urlprefix\url{https://link.aps.org/doi/10.1103/PhysRev.93.99}.

\bibitem{kirton_introduction_2019}
\bibinfo{author}{Kirton, P.}, \bibinfo{author}{Roses, M.~M.},
  \bibinfo{author}{Keeling, J.} \& \bibinfo{author}{Torre, E. G.~D.}
\newblock \bibinfo{title}{Introduction to the {{Dicke Model}}: {{From
  Equilibrium}} to {{Nonequilibrium}}, and {{Vice Versa}}}.
\newblock \emph{\bibinfo{journal}{Advanced Quantum Technologies}}
  \textbf{\bibinfo{volume}{2}}, \bibinfo{pages}{1800043}
  (\bibinfo{year}{2019}).

\bibitem{Fox}
\bibinfo{author}{Fox, M.} \& \bibinfo{author}{Javanainen, J.}
\newblock \bibinfo{title}{Quantum optics: An introduction}.
\newblock \emph{\bibinfo{journal}{Physics Today - PHYS TODAY}}
  \textbf{\bibinfo{volume}{60}} (\bibinfo{year}{2007}).

\bibitem{frisk_kockum_ultrastrong_2019}
\bibinfo{author}{Frisk~Kockum, A.}, \bibinfo{author}{Miranowicz, A.},
  \bibinfo{author}{De~Liberato, S.}, \bibinfo{author}{Savasta, S.} \&
  \bibinfo{author}{Nori, F.}
\newblock \bibinfo{title}{Ultrastrong coupling between light and matter}.
\newblock \emph{\bibinfo{journal}{Nature Reviews Physics}}
  \textbf{\bibinfo{volume}{1}}, \bibinfo{pages}{19--40} (\bibinfo{year}{2019}).
\newblock \urlprefix\url{https://www.nature.com/articles/s42254-018-0006-2}.

\bibitem{Tokatly}
\bibinfo{author}{Tokatly, I.~V.}
\newblock \bibinfo{title}{Time-dependent density functional theory for
  many-electron systems interacting with cavity photons}.
\newblock \emph{\bibinfo{journal}{Phys. Rev. Lett.}}
  \textbf{\bibinfo{volume}{110}}, \bibinfo{pages}{233001}
  (\bibinfo{year}{2013}).
\newblock
  \urlprefix\url{https://link.aps.org/doi/10.1103/PhysRevLett.110.233001}.

\bibitem{Ruggy_2014}
\bibinfo{author}{Ruggenthaler, M.} \emph{et~al.}
\newblock \bibinfo{title}{Quantum-electrodynamical density-functional theory:
  Bridging quantum optics and electronic-structure theory}.
\newblock \emph{\bibinfo{journal}{Phys. Rev. A}} \textbf{\bibinfo{volume}{90}},
  \bibinfo{pages}{012508} (\bibinfo{year}{2014}).
\newblock \urlprefix\url{https://link.aps.org/doi/10.1103/PhysRevA.90.012508}.

\bibitem{Pellegrini2015}
\bibinfo{author}{Pellegrini, C.}, \bibinfo{author}{Flick, J.},
  \bibinfo{author}{Tokatly, I.~V.}, \bibinfo{author}{Appel, H.} \&
  \bibinfo{author}{Rubio, A.}
\newblock \bibinfo{title}{Optimized effective potential for quantum
  electrodynamical time-dependent density functional theory}.
\newblock \emph{\bibinfo{journal}{Phys. Rev. Lett.}}
  \textbf{\bibinfo{volume}{115}}, \bibinfo{pages}{093001}
  (\bibinfo{year}{2015}).
\newblock
  \urlprefix\url{https://link.aps.org/doi/10.1103/PhysRevLett.115.093001}.

\bibitem{Haugland2020}
\bibinfo{author}{Haugland, T.~S.}, \bibinfo{author}{Ronca, E.},
  \bibinfo{author}{Kj\o{}nstad, E.~F.}, \bibinfo{author}{Rubio, A.} \&
  \bibinfo{author}{Koch, H.}
\newblock \bibinfo{title}{Coupled cluster theory for molecular polaritons:
  Changing ground and excited states}.
\newblock \emph{\bibinfo{journal}{Phys. Rev. X}} \textbf{\bibinfo{volume}{10}},
  \bibinfo{pages}{041043} (\bibinfo{year}{2020}).
\newblock \urlprefix\url{https://link.aps.org/doi/10.1103/PhysRevX.10.041043}.

\bibitem{Buchholz2020}
\bibinfo{author}{Buchholz, F.}, \bibinfo{author}{Theophilou, I.},
  \bibinfo{author}{Giesbertz, K. J.~H.}, \bibinfo{author}{Ruggenthaler, M.} \&
  \bibinfo{author}{Rubio, A.}
\newblock \bibinfo{title}{Light--matter hybrid-orbital-based first-principles
  methods: The influence of polariton statistics}.
\newblock \emph{\bibinfo{journal}{Journal of Chemical Theory and Computation}}
  \textbf{\bibinfo{volume}{16}}, \bibinfo{pages}{5601--5620}
  (\bibinfo{year}{2020}).
\newblock \urlprefix\url{https://doi.org/10.1021/acs.jctc.0c00469}.

\bibitem{Nielsen2018}
\bibinfo{author}{Nielsen, S. E.~B.}, \bibinfo{author}{Schäfer, C.},
  \bibinfo{author}{Ruggenthaler, M.} \& \bibinfo{author}{Rubio, A.}
\newblock \bibinfo{title}{Dressed-orbital approach to cavity quantum
  electrodynamics and beyond} (\bibinfo{year}{2018}).
\newblock \eprint{arXiv:1812.00388}.

\bibitem{Rokaj2020}
\bibinfo{author}{Rokaj, V.}, \bibinfo{author}{Ruggenthaler, M.},
  \bibinfo{author}{Eich, F.~G.} \& \bibinfo{author}{Rubio, A.}
\newblock \bibinfo{title}{Free electron gas in cavity quantum electrodynamics}.
\newblock \emph{\bibinfo{journal}{Phys. Rev. Research}}
  \textbf{\bibinfo{volume}{4}}, \bibinfo{pages}{013012} (\bibinfo{year}{2022}).
\newblock
  \urlprefix\url{https://link.aps.org/doi/10.1103/PhysRevResearch.4.013012}.

\bibitem{Li2020}
\bibinfo{author}{Li, J.} \emph{et~al.}
\newblock \bibinfo{title}{Electromagnetic coupling in tight-binding models for
  strongly correlated light and matter}.
\newblock \emph{\bibinfo{journal}{Phys. Rev. B}}
  \textbf{\bibinfo{volume}{101}}, \bibinfo{pages}{205140}
  (\bibinfo{year}{2020}).
\newblock \urlprefix\url{https://link.aps.org/doi/10.1103/PhysRevB.101.205140}.

\bibitem{sentef_quantum_2020}
\bibinfo{author}{Sentef, M.~A.}, \bibinfo{author}{Li, J.},
  \bibinfo{author}{K{\"u}nzel, F.} \& \bibinfo{author}{Eckstein, M.}
\newblock \bibinfo{title}{Quantum to classical crossover of {Floquet}
  engineering in correlated quantum systems}.
\newblock \emph{\bibinfo{journal}{Physical Review Research}}
  \textbf{\bibinfo{volume}{2}}, \bibinfo{pages}{033033} (\bibinfo{year}{2020}).
\newblock
  \urlprefix\url{https://link.aps.org/doi/10.1103/PhysRevResearch.2.033033}.

\bibitem{Dmytruk2021}
\bibinfo{author}{Dmytruk, O.} \& \bibinfo{author}{Schir\'o, M.}
\newblock \bibinfo{title}{Gauge fixing for strongly correlated electrons
  coupled to quantum light}.
\newblock \emph{\bibinfo{journal}{Phys. Rev. B}}
  \textbf{\bibinfo{volume}{103}}, \bibinfo{pages}{075131}
  (\bibinfo{year}{2021}).
\newblock \urlprefix\url{https://link.aps.org/doi/10.1103/PhysRevB.103.075131}.

\bibitem{kiffner_manipulating_2019}
\bibinfo{author}{Kiffner, M.}, \bibinfo{author}{Coulthard, J.~R.},
  \bibinfo{author}{Schlawin, F.}, \bibinfo{author}{Ardavan, A.} \&
  \bibinfo{author}{Jaksch, D.}
\newblock \bibinfo{title}{Manipulating quantum materials with quantum light}.
\newblock \emph{\bibinfo{journal}{Physical Review B}}
  \textbf{\bibinfo{volume}{99}}, \bibinfo{pages}{085116}
  (\bibinfo{year}{2019}).
\newblock \urlprefix\url{https://link.aps.org/doi/10.1103/PhysRevB.99.085116}.

\bibitem{bagchi_pedestrian_2020}
\bibinfo{author}{Bagchi, B.}, \bibinfo{author}{Ghosh, R.} \&
  \bibinfo{author}{Khare, A.}
\newblock \bibinfo{title}{A pedestrian introduction to coherent and squeezed
  states}.
\newblock \emph{\bibinfo{journal}{International Journal of Modern Physics A}}
  \textbf{\bibinfo{volume}{35}}, \bibinfo{pages}{2030011}
  (\bibinfo{year}{2020}).
\newblock \urlprefix\url{https://doi.org/10.1142/S0217751X20300112}.
\newblock \eprint{https://doi.org/10.1142/S0217751X20300112}.

\bibitem{Rabl2004}
\bibinfo{author}{Rabl, P.}, \bibinfo{author}{Shnirman, A.} \&
  \bibinfo{author}{Zoller, P.}
\newblock \bibinfo{title}{Generation of squeezed states of nanomechanical
  resonators by reservoir engineering}.
\newblock \emph{\bibinfo{journal}{Phys. Rev. B}} \textbf{\bibinfo{volume}{70}},
  \bibinfo{pages}{205304} (\bibinfo{year}{2004}).
\newblock \urlprefix\url{https://link.aps.org/doi/10.1103/PhysRevB.70.205304}.

\bibitem{Glauber1991}
\bibinfo{author}{Glauber, R.~J.} \& \bibinfo{author}{Lewenstein, M.}
\newblock \bibinfo{title}{Quantum optics of dielectric media}.
\newblock \emph{\bibinfo{journal}{Phys. Rev. A}} \textbf{\bibinfo{volume}{43}},
  \bibinfo{pages}{467--491} (\bibinfo{year}{1991}).
\newblock \urlprefix\url{https://link.aps.org/doi/10.1103/PhysRevA.43.467}.

\bibitem{Wall_optics}
\bibinfo{editor}{Walls, D.} \& \bibinfo{editor}{Milburn, G.~J.} (eds.)
  \emph{\bibinfo{title}{Quantum Optics}} (\bibinfo{publisher}{Springer Berlin
  Heidelberg}, \bibinfo{year}{2008}).
\newblock \urlprefix\url{https://doi.org/10.1007/978-3-540-28574-8}.

\bibitem{Ciuti2005}
\bibinfo{author}{Ciuti, C.}, \bibinfo{author}{Bastard, G.} \&
  \bibinfo{author}{Carusotto, I.}
\newblock \bibinfo{title}{Quantum vacuum properties of the intersubband cavity
  polariton field}.
\newblock \emph{\bibinfo{journal}{Phys. Rev. B}} \textbf{\bibinfo{volume}{72}},
  \bibinfo{pages}{115303} (\bibinfo{year}{2005}).
\newblock \urlprefix\url{https://link.aps.org/doi/10.1103/PhysRevB.72.115303}.

\bibitem{Riek420}
\bibinfo{author}{Riek, C.} \emph{et~al.}
\newblock \bibinfo{title}{Direct sampling of electric-field vacuum
  fluctuations}.
\newblock \emph{\bibinfo{journal}{Science}} \textbf{\bibinfo{volume}{350}},
  \bibinfo{pages}{420--423} (\bibinfo{year}{2015}).
\newblock \urlprefix\url{https://science.sciencemag.org/content/350/6259/420}.
\newblock
  \eprint{https://science.sciencemag.org/content/350/6259/420.full.pdf}.

\bibitem{Benea-Chelmus2019}
\bibinfo{author}{Benea-Chelmus, I.-C.}, \bibinfo{author}{Settembrini, F.~F.},
  \bibinfo{author}{Scalari, G.} \& \bibinfo{author}{Faist, J.}
\newblock \bibinfo{title}{Electric field correlation measurements on the
  electromagnetic vacuum state}.
\newblock \emph{\bibinfo{journal}{Nature}} \textbf{\bibinfo{volume}{568}},
  \bibinfo{pages}{202--206} (\bibinfo{year}{2019}).
\newblock \urlprefix\url{https://doi.org/10.1038/s41586-019-1083-9}.

\bibitem{kirton_superradiant_2018}
\bibinfo{author}{Kirton, P.} \& \bibinfo{author}{Keeling, J.}
\newblock \bibinfo{title}{{Superradiant and Lasing States in Driven-Dissipative
  {{Dicke}} Models}}.
\newblock \emph{\bibinfo{journal}{N. J. Phys.}} \textbf{\bibinfo{volume}{20}},
  \bibinfo{pages}{015009} (\bibinfo{year}{2018}).

\bibitem{Rzazewski1975}
\bibinfo{author}{Rza\ifmmode~\dot{z}\else \.{z}\fi{}ewski, K.},
  \bibinfo{author}{W\'odkiewicz, K.} \& \bibinfo{author}{\ifmmode~\dot{Z}\else
  \.{Z}\fi{}akowicz, W.}
\newblock \bibinfo{title}{Phase transitions, two-level atoms, and the ${A}^{2}$
  term}.
\newblock \emph{\bibinfo{journal}{Phys. Rev. Lett.}}
  \textbf{\bibinfo{volume}{35}}, \bibinfo{pages}{432--434}
  (\bibinfo{year}{1975}).
\newblock \urlprefix\url{https://link.aps.org/doi/10.1103/PhysRevLett.35.432}.

\bibitem{Freericks2009}
\bibinfo{author}{Freericks, J.~K.}, \bibinfo{author}{Krishnamurthy, H.~R.} \&
  \bibinfo{author}{Pruschke, T.}
\newblock \bibinfo{title}{Theoretical description of time-resolved
  photoemission spectroscopy: Application to pump-probe experiments}.
\newblock \emph{\bibinfo{journal}{Phys. Rev. Lett.}}
  \textbf{\bibinfo{volume}{102}}, \bibinfo{pages}{136401}
  (\bibinfo{year}{2009}).
\newblock
  \urlprefix\url{https://link.aps.org/doi/10.1103/PhysRevLett.102.136401}.

\bibitem{Tsuji2008}
\bibinfo{author}{Tsuji, N.}, \bibinfo{author}{Oka, T.} \&
  \bibinfo{author}{Aoki, H.}
\newblock \bibinfo{title}{Correlated electron systems periodically driven out
  of equilibrium: $\text{Floquet}+\text{DMFT}$ formalism}.
\newblock \emph{\bibinfo{journal}{Phys. Rev. B}} \textbf{\bibinfo{volume}{78}},
  \bibinfo{pages}{235124} (\bibinfo{year}{2008}).
\newblock \urlprefix\url{https://link.aps.org/doi/10.1103/PhysRevB.78.235124}.

\bibitem{Amelio2021}
\bibinfo{author}{Amelio, I.}, \bibinfo{author}{Korosec, L.},
  \bibinfo{author}{Carusotto, I.} \& \bibinfo{author}{Mazza, G.}
\newblock \bibinfo{title}{Optical dressing of the electronic response of
  two-dimensional semiconductors in quantum and classical descriptions of
  cavity electrodynamics}.
\newblock \emph{\bibinfo{journal}{Phys. Rev. B}}
  \textbf{\bibinfo{volume}{104}}, \bibinfo{pages}{235120}
  (\bibinfo{year}{2021}).
\newblock \urlprefix\url{https://link.aps.org/doi/10.1103/PhysRevB.104.235120}.

\bibitem{Scalapino_1993}
\bibinfo{author}{Scalapino, D.~J.}, \bibinfo{author}{White, S.~R.} \&
  \bibinfo{author}{Zhang, S.~C.}
\newblock \bibinfo{title}{Insulator, metal, or superconductor: The criteria}.
\newblock \emph{\bibinfo{journal}{Phys. Rev. B}} \textbf{\bibinfo{volume}{47}},
  \bibinfo{pages}{7995--8007} (\bibinfo{year}{1993}).
\newblock \urlprefix\url{https://link.aps.org/doi/10.1103/PhysRevB.47.7995}.

\bibitem{Alvermann2010}
\bibinfo{author}{Alvermann, A.}, \bibinfo{author}{Fehske, H.} \&
  \bibinfo{author}{Trugman, S.~A.}
\newblock \bibinfo{title}{Polarons and slow quantum phonons}.
\newblock \emph{\bibinfo{journal}{Phys. Rev. B}} \textbf{\bibinfo{volume}{81}},
  \bibinfo{pages}{165113} (\bibinfo{year}{2010}).
\newblock \urlprefix\url{https://link.aps.org/doi/10.1103/PhysRevB.81.165113}.

\bibitem{Giamarchi2003}
\bibinfo{author}{Giamarchi, T.}
\newblock \emph{\bibinfo{title}{Quantum Physics in One Dimension}}
  (\bibinfo{publisher}{Oxford University Press}, \bibinfo{year}{2003}).
\newblock
  \urlprefix\url{https://doi.org/10.1093/acprof:oso/9780198525004.001.0001}.

\bibitem{Gao2021}
\bibinfo{author}{Gao, H.}, \bibinfo{author}{Schlawin, F.} \&
  \bibinfo{author}{Jaksch, D.}
\newblock \bibinfo{title}{Higgs mode stabilization by photo-induced long-range
  interactions in a superconductor} (\bibinfo{year}{2021}).
\newblock \eprint{arXiv:2106.05076}.

\bibitem{katharine_collective_2020}
\bibinfo{author}{Lenk, K.} \& \bibinfo{author}{Eckstein, M.}
\newblock \bibinfo{title}{Collective excitations of the $u$(1)-symmetric
  exciton insulator in a cavity}.
\newblock \emph{\bibinfo{journal}{Phys. Rev. B}}
  \textbf{\bibinfo{volume}{102}}, \bibinfo{pages}{205129}
  (\bibinfo{year}{2020}).
\newblock \urlprefix\url{https://link.aps.org/doi/10.1103/PhysRevB.102.205129}.

\bibitem{shalabney_coherent_2015}
\bibinfo{author}{Shalabney, A.} \emph{et~al.}
\newblock \bibinfo{title}{Coherent coupling of molecular resonators with a
  microcavity mode}.
\newblock \emph{\bibinfo{journal}{Nature Communications}}
  \textbf{\bibinfo{volume}{6}}, \bibinfo{pages}{5981} (\bibinfo{year}{2015}).

\bibitem{du_can_2021}
\bibinfo{author}{Du, M.} \& \bibinfo{author}{Yuen-Zhou, J.}
\newblock \bibinfo{title}{Catalysis by dark states in vibropolaritonic
  chemistry}.
\newblock \emph{\bibinfo{journal}{Phys. Rev. Lett.}}
  \textbf{\bibinfo{volume}{128}}, \bibinfo{pages}{096001}
  (\bibinfo{year}{2022}).
\newblock
  \urlprefix\url{https://link.aps.org/doi/10.1103/PhysRevLett.128.096001}.

\bibitem{sidler_perspective_2021}
\bibinfo{author}{Sidler, D.}, \bibinfo{author}{Ruggenthaler, M.},
  \bibinfo{author}{Schäfer, C.}, \bibinfo{author}{Ronca, E.} \&
  \bibinfo{author}{Rubio, A.}
\newblock \bibinfo{title}{A perspective on ab initio modeling of polaritonic
  chemistry: {The} role of non-equilibrium effects and quantum collectivity}.
\newblock \emph{\bibinfo{journal}{arXiv:2108.12244 [physics,
  physics:quant-ph]}}  (\bibinfo{year}{2021}).
\newblock \urlprefix\url{http://arxiv.org/abs/2108.12244}.
\newblock \bibinfo{note}{ArXiv: 2108.12244}.

\bibitem{Buzzi2020}
\bibinfo{author}{Buzzi, M.} \emph{et~al.}
\newblock \bibinfo{title}{Photomolecular high-temperature superconductivity}.
\newblock \emph{\bibinfo{journal}{Phys. Rev. X}} \textbf{\bibinfo{volume}{10}},
  \bibinfo{pages}{031028} (\bibinfo{year}{2020}).
\newblock \urlprefix\url{https://link.aps.org/doi/10.1103/PhysRevX.10.031028}.

\bibitem{RevModPhys.77.513}
\bibinfo{author}{Braunstein, S.~L.} \& \bibinfo{author}{van Loock, P.}
\newblock \bibinfo{title}{Quantum information with continuous variables}.
\newblock \emph{\bibinfo{journal}{Rev. Mod. Phys.}}
  \textbf{\bibinfo{volume}{77}}, \bibinfo{pages}{513--577}
  (\bibinfo{year}{2005}).
\newblock \urlprefix\url{https://link.aps.org/doi/10.1103/RevModPhys.77.513}.

\bibitem{zhang_collective_2016}
\bibinfo{author}{Zhang, Q.} \emph{et~al.}
\newblock \bibinfo{title}{Collective non-perturbative coupling of {2D}
  electrons with high-quality-factor terahertz cavity photons}.
\newblock \emph{\bibinfo{journal}{Nature Phys}} \textbf{\bibinfo{volume}{12}},
  \bibinfo{pages}{1005--1011} (\bibinfo{year}{2016}).
\newblock \urlprefix\url{https://www.nature.com/articles/nphys3850}.
\newblock \bibinfo{note}{Bandiera\_abtest: a Cg\_type: Nature Research Journals
  Number: 11 Primary\_atype: Research Publisher: Nature Publishing Group
  Subject\_term: Quantum Hall;Quantum optics Subject\_term\_id:
  quantum-hall;quantum-optics}.

\bibitem{ruggenthaler_quantum-electrodynamical_2018}
\bibinfo{author}{Ruggenthaler, M.}, \bibinfo{author}{Tancogne-Dejean, N.},
  \bibinfo{author}{Flick, J.}, \bibinfo{author}{Appel, H.} \&
  \bibinfo{author}{Rubio, A.}
\newblock \bibinfo{title}{From a quantum-electrodynamical light–matter
  description to novel spectroscopies}.
\newblock \emph{\bibinfo{journal}{Nat Rev Chem}} \textbf{\bibinfo{volume}{2}},
  \bibinfo{pages}{1--16} (\bibinfo{year}{2018}).
\newblock \urlprefix\url{https://www.nature.com/articles/s41570-018-0118}.
\newblock \bibinfo{note}{Bandiera\_abtest: a Cg\_type: Nature Research Journals
  Number: 3 Primary\_atype: Reviews Publisher: Nature Publishing Group
  Subject\_term: Chemical physics;Method development;Quantum chemistry;Quantum
  physics Subject\_term\_id:
  chemical-physics;method-development;quantum-chemistry;quantum-physics}.

\bibitem{truax_baker-campbell-hausdorff_1985}
\bibinfo{author}{Truax, D.~R.}
\newblock \bibinfo{title}{Baker-{Campbell}-{Hausdorff} relations and unitarity
  of {SU}(2) and {SU}(1,1) squeeze operators}.
\newblock \emph{\bibinfo{journal}{Phys. Rev. D}} \textbf{\bibinfo{volume}{31}},
  \bibinfo{pages}{1988--1991} (\bibinfo{year}{1985}).
\newblock \urlprefix\url{https://link.aps.org/doi/10.1103/PhysRevD.31.1988}.

\bibitem{van-brunt_special-case_2015}
\bibinfo{author}{Van-Brunt, A.} \& \bibinfo{author}{Visser, M.}
\newblock \bibinfo{title}{Special-case closed form of the
  {Baker}-{Campbell}-{Hausdorff} formula}.
\newblock \emph{\bibinfo{journal}{J. Phys. A: Math. Theor.}}
  \textbf{\bibinfo{volume}{48}}, \bibinfo{pages}{225207}
  (\bibinfo{year}{2015}).
\newblock \urlprefix\url{http://arxiv.org/abs/1501.02506}.
\newblock \bibinfo{note}{ArXiv: 1501.02506}.

\bibitem{Mahan1990}
\bibinfo{author}{Mahan, G.~D.}
\newblock \emph{\bibinfo{title}{Many-Particle Physics}}
  (\bibinfo{publisher}{Springer {US}}, \bibinfo{year}{1990}).
\newblock \urlprefix\url{https://doi.org/10.1007/978-1-4613-1469-1}.

\end{thebibliography}

\section*{ACKNOWLEDGEMENT} 

The authors thank Vasilis Rokaj, Brieuc Le De, Martin Eckstein, Jiajun Li and Mara Caltapanides for fruitful discussions regarding the manuscript.

We acknowledge support by the Deutsche Forschungsgemeinschaft (DFG, German Research Foundation) via  Germany’s Excellence Strategy -- Cluster of Excellence Matter and Light for Quantum Computing (ML4Q) EXC 2004/1 -- 390534769 and within the RTG 1995. We also acknowledge support from the Max Planck-New York City Center for Non-Equilibrium Quantum Phenomena. MAS acknowledges financial support through the Deutsche Forschungsgemeinschaft (DFG, German Research Foundation) via the Emmy Noether program (SE 2558/2). C.K. acknowledges support by the Deutsche Forschungsgemeinschaft (DFG, German Research Foundation) through the Emmy Noether program (KA3360/2-1) as well as by `Nieders\"achsisches Vorab' through the `Quantum- and Nano-Metrology (QUANOMET)' initiative  within  the  project  P-1. M.O. gratefully acknowledges the support of the Braunschweig International Graduate School of Metrology B-IGSM and the DFG Research Training Group 1952 Metrology for Complex Nanosystems.

\section*{Author contributions}
C.J.E. carried out the simulations with the variational code, G.P. and M.O. performed the ED simulations.
Analytical calculations were done by C.J.E.
All authors analyzed the data and discussed the results.
C.J.E., G.P., M.A.S. and D.M.K. wrote the manuscript with input from M.O., F.C. and C.K.
The project was conceived by D.M.K. and M.A.S.

\section*{Competing Interests}
The authors declare no competing interests.

\appendix

\section*{{\bf Supplementary information}}
\section*{Supplementary Note 1: Collective strong coupling in the case of $N$ identical modes}
In this supplementary we show that coupling electrons to $N$ identical modes with a coupling constant $\frac{g}{\sqrt{L}}$, in a setup as described in the Model subsection under Results of the main text, effectively results in a single mode coupling with enhanced strength $\frac{g \, \sqrt{N}}{\sqrt{L}}$ and $N-1$ completely decoupled modes. We will start to show how this holds for the Hamiltonian expanded to second order in the light-matter coupling (also compare Eq.~(15) of the main text) and later in this section argue why this might also hold for the full Peierls substitution including all order in the LMC (also compare Eq.~(2) of the main text).
We write the Hamiltonian to second order in the LMC $g$ for many identical modes as
\begin{equation}
    H = \mathcal{T} + \frac{g}{\sqrt{L}} \mathcal{J} \sum_{\lambda} \left(a_{\lambda}^{\dag} + a_{\lambda}\right)  - \frac{1}{2} \frac{g^2}{L} \mathcal{T} \left(\sum_{\lambda} \left(a_{\lambda}^{\dag} + a_{\lambda}\right) \right)^2 + \omega \sum_{\lambda} a_{\lambda}^{\dag} a_{\lambda}.
\end{equation}
Here $a_{\lambda}$ annihilates -; $a_{\lambda}^{\dag}$ creates a photon in mode $\lambda$.
All other symbols are as defined in the main text.

To find a form where the modes are decoupled we will represent them in terms of their generalized coordinate and momentum according to
\begin{equation}
    \begin{aligned}
    X_{\lambda} &= \frac{1}{\sqrt{2 \omega}} \left(a_{\lambda}^{\dag} + a_{\lambda}\right)\\
    P_{\lambda} &= i\frac{\sqrt{\omega}}{\sqrt{2}} \left(a_{\lambda}^{\dag} - a_{\lambda}\right)
    \label{eq:positionAndMomentum}
    \end{aligned}
\end{equation}
with which the Hamiltonian becomes
\begin{equation}
    H = \mathcal{T} + \sqrt{2 \omega} \frac{g}{\sqrt{L}} \mathcal{J} \sum_{\lambda} X_{\lambda} - \frac{g^2}{L} \omega \mathcal{T} \sum_{\lambda, \kappa} X_{\lambda} X_{\kappa} + 
    \sum_{\lambda} \frac{1}{2} \omega^2 X_{\lambda}^2 + \frac{1}{2} P_{\lambda}^2.
\end{equation}
This can be written in matrix form as
\begin{equation}
    H = \mathcal{T} + \sqrt{2 \omega} \frac{g}{\sqrt{L}} \mathcal{J} \sum_{\lambda} X_{\lambda} - \frac{g^2}{L} \omega \mathcal{T} %
    \, \underline{X}^{\mathrm{T}}%
    \begin{pmatrix}
    I_{\mathrm{e}} - \frac{\omega \, L}{2 g^2} \mathcal{T}^{-1} & I_{\mathrm{e}} & \dots & I_{\mathrm{e}}\\
    I_{\mathrm{e}} & I_{\mathrm{e}} - \frac{\omega \, L}{2 g^2} \mathcal{T}^{-1} & I_{\mathrm{e}} & \dots\\
    \dots & & \dots & I_{\mathrm{e}} &\\
    I_{\mathrm{e}} & \dots & I_{\mathrm{e}} & I_{\mathrm{e}} - \frac{\omega \, L}{2 g^2} \mathcal{T}^{-1}
    \end{pmatrix}%
    \underline{X}%
    + \frac{1}{2} %
    \underline{P}^{\mathrm{T}}
    \, I_{N {\times} N} \,%
    \underline{P}.
\label{eq:manymodes_HamiltonianMatrix}
\end{equation}
Here $I_{\mathrm{e}}$ is the identity on the electronic part of the Hilbert space.
We have introduced $N$-dimensional coordinate and momentum vectors as
\begin{equation}
    \underline{X} =%
    \begin{pmatrix}
    X_1 \\
    \dots \\
    X_N
    \end{pmatrix}%
    \hspace{2mm} ; \hspace{3mm}
        \underline{P} =%
    \begin{pmatrix}
    P_1 \\
    \dots \\
    P_N
    \end{pmatrix}%
    \label{eq:NDim}
\end{equation}
and $I_{N{\times}N}$ is simply the unity in $N$ dimensions with $I_{\mathrm{e}}$ on the diagonal.
One eigenvector of the above matrix in Eq.~(\ref{eq:manymodes_HamiltonianMatrix}) is clearly 
\begin{equation}
    v^1 = \frac{1}{\sqrt{N}}%
    \begin{pmatrix}
    1\\
    \dots \\
    1
    \end{pmatrix}
\end{equation}
with corresponding eigenvalue (that still contains an operator from the electronic subsystem due to the composite nature of the system)
\begin{equation}
    \varepsilon^1 = N - \frac{\omega \, L}{2 g^2} \mathcal{T}^{-1}.
\end{equation}
Each vector $v = (v_1, \dots, v_N)^{\mathrm{T}}$ from the orthogonal $N-1$ dimensional subspace of $v^1$, defined through the equation $\sum_{i = 1}^{N} v_i = 0$, is an eigenvector with eigenvalue $\varepsilon = - \frac{\omega \, L}{2 g^2} \mathcal{T}^{-1}$ which is therefore $N-1$ times degenerate.
Denoting by $P_+$ and $X_+$ momentum and coordinate corresponding to the first eigenvector and by $\tilde{P}_{\kappa}$, $\Tilde{X}_{\kappa}$, $\kappa = 1, \dots, N-1$ momenta and coordinates corresponding to the other $N-1$ eigenvectors
we can write the Hamiltonian with decoupled bosonic modes as
\begin{equation}
    H = \mathcal{T} + \sqrt{2 \omega} \sqrt{N} \frac{g}{\sqrt{L}} X_+ \mathcal{J} + \frac{1}{2} \left( \omega^2 - 2N \frac{g^2}{L} \omega \mathcal{T}\right) X_+^2 + \frac{1}{2} P_+^2%
    + \sum_{\kappa} \frac{1}{2} \omega^2 \Tilde{X}_{\kappa}^2 + \frac{1}{2} \Tilde{P}_{\kappa}^2.
    \label{eq:supplementaryDecoupling_decoupledHamiltonian}
\end{equation}
From this it is clear that the $X_+$ mode couples to the electrons with effective strength $\frac{g \sqrt{N}}{\sqrt{L}}$ while all other $N - 1$ modes don't couple to the electrons or among each other at all.

Next we discuss the case of the full Peierls substitution keeping all orders in the LMC.
For this situation the Hamiltonian including many identical modes with zero momentum transfer would read
\begin{equation}
    H = \sin \left(\frac{g}{\sqrt{L}} \sum_{\lambda} \left(a_{\lambda}^{\dag} + a_{\lambda}\right)\right)  \mathcal{J}  +  \cos \left(\frac{g}{\sqrt{L}} \sum_{\lambda} \left(a_{\lambda}^{\dag} + a_{\lambda}\right) \right) \mathcal{T} + \omega \sum_{\lambda} a_{\lambda}^{\dag} a_{\lambda}.
\end{equation}
We now write this Hamiltonian in terms of the canonical position and momentum operators introduced in Eq.~(\ref{eq:positionAndMomentum})
\begin{equation}
\begin{aligned}
    H &= \sin \left(\frac{g\sqrt{2\omega}}{\sqrt{L}} \sum_{\lambda} X_{\lambda}\right)  \mathcal{J}  +  \cos \left(\frac{g \sqrt{2\omega}}{\sqrt{L}} \sum_{\lambda} X_{\lambda} \right) \mathcal{T} + \sum_{\lambda} \frac{1}{2} P_{\lambda}^2 + \frac{\omega^2}{2} X_{\lambda}^{2}\\
    & = \sin \left(\frac{g\sqrt{2\omega}}{\sqrt{L}} \sum_{\lambda} X_{\lambda}\right)  \mathcal{J}  +  \cos \left(\frac{g \sqrt{2\omega}}{\sqrt{L}} \sum_{\lambda} X_{\lambda} \right) \mathcal{T} + 
    \frac{1}{2} %
    \underline{P}^{\mathrm{T}}
    \, I_{N {\times} N} \,%
    \underline{P} + 
    \frac{\omega^2}{2} %
    \underline{X}^{\mathrm{T}}
    \, I_{N {\times} N} \,%
    \underline{X}
\end{aligned}
\end{equation}
where in the last step we have again introduced $N$-dimensional notation as in Eq.~(\ref{eq:NDim}).
The fact that the harmonic oscillator terms can be written using the $N$-dimensional unity $I_{N \times N}$ stems from our approximation of all modes having equal frequency.
Due to this, we can now write the Hamiltonian in terms of any other set of collective modes in particular the one used to write Eq.~(\ref{eq:supplementaryDecoupling_decoupledHamiltonian}) in which the last term will remain diagonal (ie. in particular not couple different modes) obtaining
\begin{equation}
\begin{aligned}
    H &= \sin \left(\frac{g\sqrt{2\omega} \sqrt{N}}{\sqrt{L}} X_{+}\right)  \mathcal{J}  +  \cos \left(\frac{g \sqrt{2\omega} \sqrt{N}}{\sqrt{L}} X_{+} \right) \mathcal{T} + \frac{1}{2} P_{+}^2 + \frac{\omega^2}{2} X_{+}^{2} + \sum_{\kappa = 1}^{N - 1} \frac{1}{2} \Tilde{P}_{\kappa}^2 + \frac{\omega^2}{2} \Tilde{X}_{\kappa}^{2}.
    \label{eq:supplementaryDecoupling_decoupledHamiltonianAllOrders}
\end{aligned}
\end{equation}
Here all operators are defined as in Eq.~(\ref{eq:supplementaryDecoupling_decoupledHamiltonian}).
Thus also in the case of keeping all orders in the LMC we obtain a single mode with effectively enhanced coupling $\frac{g}{\sqrt{L}} \rightarrow \frac{g \sqrt{N}}{\sqrt{L}}$ and $N - 1$ uncoupled modes.


In Eq.~(\ref{eq:supplementaryDecoupling_decoupledHamiltonian}) (and also Eq.~(\ref{eq:supplementaryDecoupling_decoupledHamiltonianAllOrders}) when expanding again) it seems like the effective frequency of the $X_+$ mode would scale like $\sqrt{N}$ for large enough $N$ which seems counter-intuitive.
This is, however, reminiscent of the dipole approximation that is here taken for all modes.
When allowing for any small but non-zero momentum transfer, the modes immediately couple to a microscopic quantity instead of all electrons collectively yielding a finite effective frequency. 

The here shown mechanism for collective strong coupling is reminiscent of an analogous one considered in the case of vibrational Strong Coupling\cite{shalabney_coherent_2015, du_can_2021, sidler_perspective_2021} in the case of a cavity coupling to vibrational excitations of a solid or to collective strong coupling of an electro-magnetic resonator coupled to many emitters.\cite{frisk_kockum_ultrastrong_2019, sidler_perspective_2021}

\section*{Supplementary Note 2: Diagonalization of the Hamiltonian in the TD limit}\label{sec:supp_ground}

In this part we show how to diagonalize the Hamiltonian expanded to second order in the field that gave the only non-vanishing contribution in the TD limit to the GS energy in Eq.~(6). It reads
\begin{equation}
    \tag{S1}
    H^{2^{\mathrm{nd}}} = \omega_0 \left(a^{\dag}a + \frac{1}{2}\right) + %
\mathcal{T} + %
\frac{g}{\sqrt{L}} \left(a^{\dagger} + a\right) \mathcal{J} %
- \frac{g^2}{2 L} (\aplusa)^2 \mathcal{T}
\end{equation}
and can be diagonalized using a combined squeezing and displacement transformation\cite{Kennes2017, bagchi_pedestrian_2020}
\begin{equation}
\tag{S2}
\begin{aligned}
    H^{\mathrm{D}} &= e^{S^{\mathrm{d}}[\mathcal{T}, \mathcal{J}]} e^{S^{\mathrm{sq}}[\mathcal{T}]} H^{A, A^2} e^{-S^{\mathrm{d}}[\mathcal{T}, \mathcal{J}]} e^{-S^{\mathrm{sq}}[\mathcal{T}]}\\
    S^{\text{d}}[\mathcal{T}, \mathcal{J}] &= \frac{g}{\sqrt{L} \omega_0 } \left( \frac{\mathcal{W}[\mathcal{T}]}{\omega_0} \right)^{-\frac{3}{2}}%
\left(a^{\dag} - a\right) \mathcal{J},\\
    S^{\text{sq}}[\mathcal{T}] &= \frac{1}{4} \ln \left( \frac{\mathcal{W}[\mathcal{T}]}{\omega_0} \right) \left(a^2 - (a^{\dagger})^2\right).
     \end{aligned}
     \label{eq:squeezingAndDisplacement}
\end{equation}
The diagonal Hamiltonian $H^{\mathrm{D}}$ is given in the main text Eq.~(7) together with the definition of $\mathcal{W}[\mathcal{T}]$.
Both displacement and squeezing transformations depend on fermionic operators namely the kinetic energy $\mathcal{T}$ and the current $\mathcal{J}$.
Since $\mathcal{T}$ and $\mathcal{J}$ are diagonal in $k$-space the GS of the whole system is given as (see also Eq.~(11) of the main text and below)
\begin{equation}
\tag{S3}
\begin{aligned}
       | \Phi_{\mathrm{GS}} \rangle &= |\psi_{\mathrm{GS}}\rangle_f \otimes |0_{\beta} \rangle\\
       &= |\psi_{\mathrm{GS}}\rangle_f \otimes e^{S^{\mathrm{d}}[-t_{\mathrm{GS}}L, j_{\mathrm{GS}}L]} e^{S^{\mathrm{sq}}[-t_{\mathrm{GS}}L]} |0\rangle.
\end{aligned}
\end{equation}
where $|\psi_{\mathrm{GS}}\rangle_f$ is the unshifted FS and $|0_{\beta} \rangle$ is the vacuum state of the annihilators(creators) $\beta^{(\dagger)}$ of the coherent squeezed states, defined in the main text Eq.~(8).
$|0\rangle$ is the vacuum state of the non squeezed bosonic operators $a^{\dag}$ and $a$.
Since we found $j_{\mathrm{GS}} = 0$ due to the vanishing shift of the FS we have $e^{S^{\mathrm{d}}[-t_{\mathrm{GS}}L, j_{\mathrm{GS}}L]_{\beta}} = I_b$ where $I_b$ is the identity on the bosonic part of the Hilbertspace.
The photon part of the GS wavefunction is thus given by Eq.~(12) of the main part.

\section*{Supplementary Note 3: Momentum-resolved spectral function in the TD limit}
\label{sec:AppSpectralFunction}%
In this part we show how to analytically calculate the spectral function $A(k, \omega)$ of the electrons in the TD limit.
Since we do this at temperature $T=0$ the expectation values appearing in the definition of the spectral function (Eq.~(16) of the main text) are taken just with respect to the GS.
None of the operators in the expectation value creates a macroscopic occupation of the photonic mode.
Therefore, the scaling analysis of Eq.~(6) of the main text can be applied in this case allowing us to diagonalize the problem by the combined squeezing and displacement transformation Eq.~(\ref{eq:squeezingAndDisplacement}).
To evaluate the expectation values we also need the behaviour of the fermionic creation (annihilation) operators under these transformations which read
\begin{equation}
\tag{S6}
\begin{aligned}
& e^{S^{\text{d}}} e^{S^{\text{sq}}} c_k e^{-S^{\text{sq}}} e^{-S^{\text{d}}} = c_k X Y,\\
& e^{S^{\text{d}}} e^{S^{\text{sq}}} c_k^{\dag} e^{-S^{\text{sq}}} e^{-S^{\text{d}}} = %
c_k^{\dag} X^{\dag} Y^{\dag}
\end{aligned}
\end{equation}
with
\begin{equation}
\tag{S7}
\begin{aligned}
\ln (X) = - \frac{g \omega_0 \mathcal{W}^{-2}}{\sqrt{L}} v_k \left(a^{\dag} - a\right) + \mathcal{O}\left(\frac{1}{L^{\frac{3}{2}}}\right),\\
\ln (Y) = \frac{1}{2} \frac{g^2}{\omega_0 L} \varepsilon_k \left(1 - 2 \frac{g^2}{\omega_0 L} \mathcal{T}\right)^{-1} \left(a^2 - (a^{\dag})^2\right) + \mathcal{O}\left(\frac{1}{L^{\frac{3}{2}}}\right).
\end{aligned}
\end{equation}
Considering the first expectation value from the spectral function, Eq.~(16) of the main text, we find
\begin{equation}
\tag{S8}
\begin{aligned}
\langle c_k(t) c_k^{\dag} \rangle &= \, _f\bra{\psi_{\mathrm{GS}}} \otimes \, _b\bra{\phi_{\mathrm{GS}}} %
\overbrace{1}^{\mathclap{e^{-S^{\text{sq}}[\mathcal{T}]} e^{-S^{\text{d}}[\mathcal{T}, \mathcal{J}]} e^{S^{\text{d}}[\mathcal{T}, \mathcal{J}]} e^{S^{\text{sq}}[\mathcal{T}]}}}%
e^{i H t} c_k e^{- i H t} c_k^{\dag} %
\underbrace{1}_{\mathclap{e^{-S^{\text{sq}}[\mathcal{T}]} e^{-S^{\text{d}}[\mathcal{T}, \mathcal{J}]} e^{S^{\text{d}}[\mathcal{T}, \mathcal{J}]} e^{S^{\text{sq}}[\mathcal{T}]}}}%
\ket{\phi_{\mathrm{GS}}}_b \otimes \ket{\psi_{\mathrm{GS}}}_f \\
& = \, _f\bra{\psi_{\mathrm{GS}}} \otimes \bra{0} %
e^{i H^{\text{D}}t} c_k X Y e^{-i H^{\text{D}}t}%
c_k^{\dag} X^{\dag} Y^{\dag}%
\ket{0} \otimes \ket{\psi_{\mathrm{GS}}}_f + \mathcal{O}\left(\frac{1}{L^{\frac{3}{2}}}\right)\\
& = \bra{\psi_{\mathrm{GS}}}_f \otimes \bra{0} %
c_k(t)_{H^{\text{D}}} X(t)_{H^{\text{D}}} Y(t)_{H^{\text{D}}}%
c_k^{\dag} X^{\dag} Y^{\dag}%
\ket{0} \otimes \ket{\psi_{\mathrm{GS}}}_f  + \mathcal{O}\left(\frac{1}{L^{\frac{3}{2}}}\right).
\end{aligned} 
\label{eq:squeezedExpectationValue}
\end{equation}
With the subscript $(.)(t)_{H^{\text{D}}}$ we signify that the time dependence is determined by the diagonal Hamiltonian $H^{\text{D}}$, Eq.~(7) of the main text.

The operators $\mathcal{T}$ and $\mathcal{J}$ appearing in $X$ and $Y$ have no time dependence since they commute with $H^{\text{D}}$ (and in fact also the full $H$).
The time dependence of the operators $X$ and $Y$ is determined by that of the bosonic operators
\begin{equation}
\tag{S9}
\begin{aligned}
a(t)_{H^{\text{D}}} & = a e^{- i \mathcal{W} t}\\
a^{\dag}(t)_{H^{\text{D}}} &= a^{\dag} e^{i \mathcal{W} t}.
\end{aligned}
\end{equation}
Evaluating the electronic part of the expectation value will yield $\mathcal{W} \to \tilde{\omega}$ restoring a simple time dependence with the dressed cavity frequency $\tilde{\omega}$.

Reconsidering the expectation value Eq.~(\ref{eq:squeezedExpectationValue}) we note that moving the fermionic operators through $X$ and $Y$ will only yield higher order corrections such that we can write
\begin{equation}
\tag{S10}
\langle c_k(t) c_k^{\dag} \rangle = e^{\Phi(t)} (1 - n_k) \bra{0} %
X_{\psi_{\text{GS}}}(t)_{H^{\text{D}}_b} Y_{\psi_{\text{GS}}}(t)_{H^{\text{D}}_b} X^{\dag}_{\psi_{\text{GS}}} Y^{\dag}_{\psi_{\text{GS}}} \ket{0}
\label{eq:fermionicEvaluated}
\end{equation}
where $n_k = \langle c_k^{\dag}c_k \rangle$.
Here we have evaluated the time dependence of the fermionic annihilators that yields the time dependent phase factor $e^{\Phi(t)}$. We find, only keeping the leading order as before 
\begin{equation}
\tag{S11}
c_k(t)_{H^{\text{D}}} = c_k e^{\mathcal{F}(t)} \hspace{2mm}; \hspace{2mm}
\mathcal{F}(t) = -i \varepsilon_k t +  %
i \frac{g^2 \varepsilon_k}{L} \omega_0 \mathcal{W}^{-1} \left(\ada + \frac{1}{2}\right)t%
- i \frac{g^2 \omega_0 \mathcal{W}^{-2}}{L} v_k^2 t
\end{equation}
Evaluating the expectation of this yields
\begin{equation}
\tag{S12}
    \langle e^{\mathcal{F}(t)} \rangle = e^{\Phi(t)}  \hspace{2mm}; \hspace{2mm}
\Phi(t) = -i \varepsilon_k t +  %
i \frac{g^2 \varepsilon_k}{2L} \frac{\omega_0}{\tilde{\omega}} t%
- i \Sigma_k t
\end{equation}
with 
\begin{equation}
\tag{S13}
    \Sigma_k = \frac{g^2 \omega_0}{\tilde{\omega}^2 L} v_k^2.
\end{equation}
In Eq.~(\ref{eq:fermionicEvaluated}) we have already executed the fermionic part of the expectation value performing
\begin{equation}
\tag{S14}
\begin{aligned}
\frac{\mathcal{T}}{L} &\rightarrow \frac{\bra{\psi_{\mathrm{GS}}}_f \mathcal{T} \ket{\psi_{\mathrm{GS}}}_f}{L} = t_{\text{GS}}\\
\frac{\mathcal{J}}{L} &\rightarrow \frac{\bra{\psi_{\mathrm{GS}}}_f \mathcal{J} \ket{\psi_{\mathrm{GS}}}_f}{L} = j_{\text{GS}}
\end{aligned}
\end{equation}
in the $X^{(\dag)}$ and $Y^{(\dag)}$ operator writing them as $X^{(\dag)}_{\psi_{\text{GS}}}$ and $Y^{(\dag)}_{\psi_{\text{GS}}}$.

Since all operators act on the $\ket{0}$ state, contributions come only from commutators of the operators in the exponentials.
Therefore, all contributions from the $Y$ operator are at least $\exp\left(\mathcal{O}\left(\frac{1}{L^{\frac{3}{2}}}\right)\right)$ \cite{truax_baker-campbell-hausdorff_1985, van-brunt_special-case_2015} and will thus be neglected.
We are thus left with
\begin{equation}
\tag{S15}
\langle c_k(t) c_k^{\dag} \rangle = e^{\Phi(t)} (1 - n_k) \bra{0} X_{\psi_{\text{GS}}}(t)_{H^{\mathrm{D}}_b} X^{\dag}_{\psi_{\text{GS}}} \ket{0} + \mathcal{O}\left( \frac{1}{L^{\frac{3}{2}}} \right).
\label{eq:finalExpectation}
\end{equation}
The evaluation of the remaining expectation value is a standard textbook problem. \cite{Mahan1990}

Evaluating the other expectation value in the definition of the spectral function (Eq.~(16) in the main part) yields the same result, just with a factor $n_k$ instead of $1 - n_k$ up front and the final expectation value is in Eq.~(\ref{eq:finalExpectation}) is complex conjugated as the order of the operators is reversed.
This reflects the particle-hole symmetry of the half-filled system, which is inherited from the bare chain. 

Performing the remaining FT we arrive at the final result reported in Eq.~(18) in the main text.

\section*{Supplementary Note 4: Non-equilibrium spectral function from coherent pumping}
\begin{figure*}[t]
\centering
    \includegraphics[scale = 1.0]{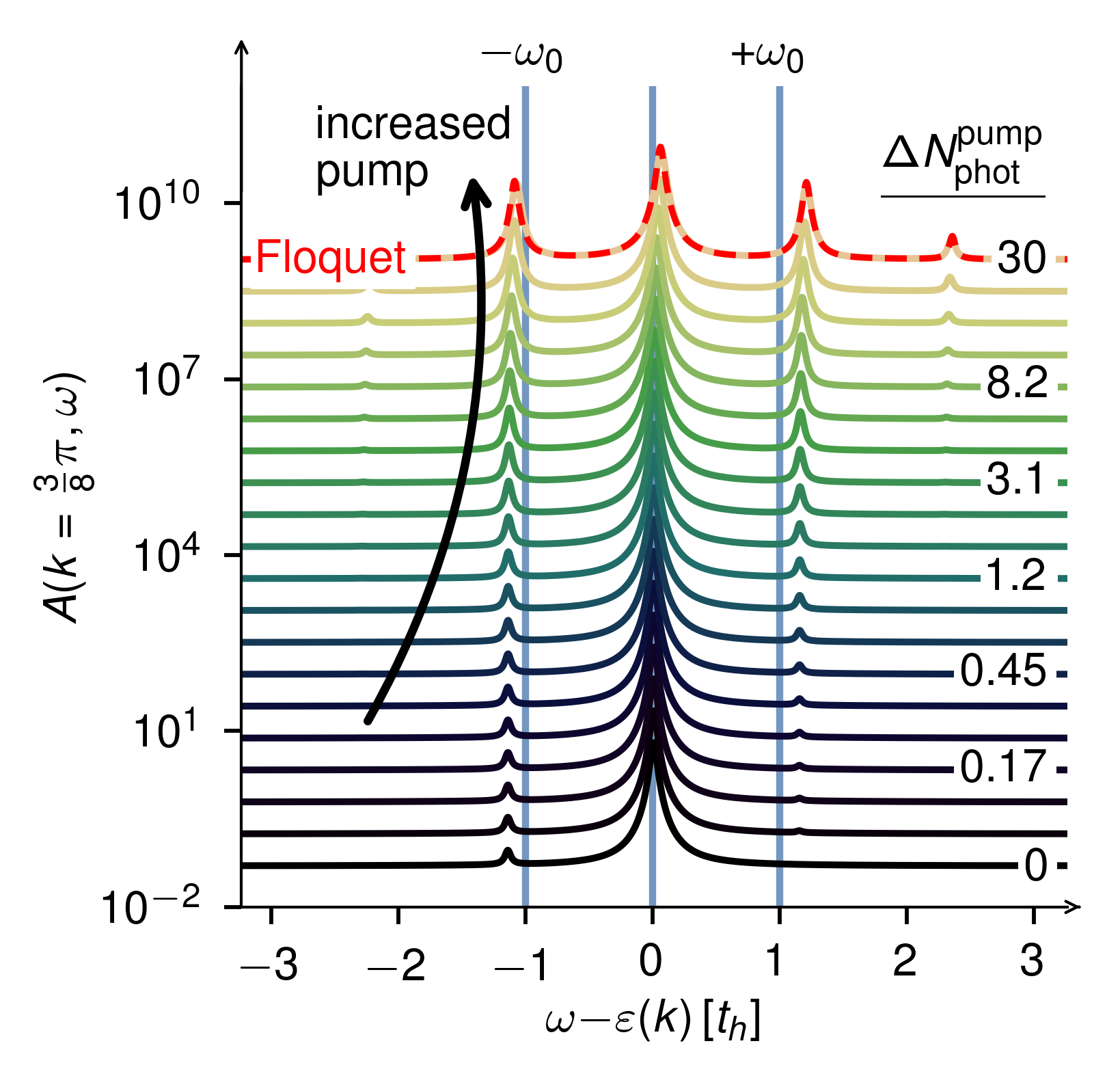}
    
    \caption{\textbf{Strong pumping limit of the non-equilibrium spectral function.}
    Non-equilibrium spectral function obtained according to Eq.~(22) in analogy to Fig.~3(b) of the main text.
    The LMC is kept constant at $g = 0.5$ while the strength of the pump increases from zero to $\Delta N_{\rm phot}^{\rm pump} = 30$ as reported on the right-hand side of the plot.
    The spectral function corresponding to the strongest pump is overlayed with the non-equilibrium spectral function of the classically driven system (Eq.~(24)) at the effective cavity frequency $\tilde{\omega}$ as stated in Eq.~(10).
    In analogy to Fig.~3(b) of the main text, the structure of the peaks changes from completely asymmetric to symmetric for increased pumping.
    In contrast to Fig.~3(b), the size of the side-peaks now increases for stronger pumping. Additionally, features that were previously small in the TD limit (see Eq.~(18)) now emerge as for example the dynamical localization (shift of central peak) and the shake-off bands (also a second shake-off band is now visible).
    These features are well reproduced within the classical drive.
    Parameters, if not specifically mentioned otherwise, are as in Fig.~3(b) but with an increased size of the bosonic Hilbertspace of $N_{\rm max}^{\rm boson} = 130$.}
\end{figure*}
In this part we calculate the non-equilibrium spectral function according to Eq.~(22) of the main text in analogy to our analysis in the Quantum to Floquet crossover subsection under Results in the main text.
However, in contrast to that part, we do not keep $g^2 \, \Delta N_{\rm phot}^{\rm pump} = \rm const$ while sending $\Delta N_{\rm phot}^{\rm pump} \to \infty$ but set $g = 0.5$.
Hence, we here do not perform the classical limit, since the light-matter hybridization is never lifted, but the limit of strong driving.

The result can be seen in Fig.~1 of this supplement.
The side-peaks are at a shifted frequency $\tilde{\omega}$ which reflects the fact that the effective boson of the system represents a mixture of light and matter degrees of freedom. 
In contrast to the classical limit, their position stays approximately constant and does not reduce to $\omega_0$ for stronger pumping.
At the same time, the evolution of completely asymmetric side-peaks to fully symmetric ones prevails.
The strength of the peaks increases monotonically with stronger pumping while it stayed almost constant previously.

The last line corresponding to the strongest pump is again compared to the non-equilibrium spectral function obtained from a classically driven system according to Eq.(24) of the main text.
We set the frequency to $\tilde{\omega}$ as stated in Eq.~(10).
The result matches well with that of the strongest drive.
For even larger numbers of photons injected into the system one will, however, start to see deviations as the higher, non-harmonic terms in the Hamiltonian become relevant for the dynamics.

As expected, features of the electronic spectral function that were previously small in the TD limit (see Eq.~(18) of the main text) are enhanced through the driving.
The dynamical localization now becomes notable through the shift of the central peak and a second shake-off band appears.
These features are also well reproduced by the classical drive in this regime.


\end{document}